\documentclass[12pt,a4paper]{article}
\pdfoutput=1
\usepackage{jheppub}
\usepackage[utf8]{inputenc}
\usepackage[T1]{fontenc}
\usepackage{comment}

\usepackage{amsmath,physics,float,tensor}  
\usepackage{amssymb}  
\usepackage{latexsym}
\usepackage{graphicx,xcolor}

\usepackage{amsfonts}
\usepackage{braket}
\usepackage{mathrsfs}

\usepackage{subfig} 
\usepackage{empheq}

\usepackage[shortlabels]{enumitem}

\usepackage{pgfplots}
\usepackage{natbib}

\usepackage{bigints}

\newcommand{\beq}{\begin{equation}}
\newcommand{\eeq}{\end{equation}}

\newcommand{\del}{\partial}
\newcommand{\lc}{\left(}
\newcommand{\rc}{\right)}
\newcommand{\ls}{\left[}
\newcommand{\rs}{\right]} 
\newcommand{\pdel}{\mathcal{P}_\delta}
\newcommand{\ms}{m}%scalar mass
\newcommand{\mg}{\mathfrak{m}}  % spin2 mass
\newcommand{\ps}{\pi_{\phi}}%scalar momenta

\newcommand{\tc}[2]{\textcolor{1}{2}}
\newcommand{\pd}{\partial}

\newcommand{\bPB}[2]{\left\{#1,#2\right\}}

\newcommand{\D}{\text{D}}

\newcommand{\Csn}{
C}  %full constraints redefined
\newcommand{\grf}{h}  %spin2 field
\newcommand{\Pg}{\Pi} % spin2 conj. momentum
\newcommand{\tot}{\text{tot}}

\usepackage{tikz, pgf}
\usetikzlibrary{mindmap} 
\begin{document}

\title{\boldmath Holography of information in massive gravity using Dirac brackets}
\author[1,2]{Joydeep Chakravarty,}
\emailAdd{joydeep.chakravarty@mail.mcgill.ca}

\author[3]{Diksha Jain,}
\emailAdd{diksha.2012jain@gmail.com}

\author[2,4]{Akhil Sivakumar}
\emailAdd{akhil.sivakumar@apctp.org}

\affiliation[1]{Department of Physics, McGill University,\\
3600 Rue University, Montreal, H3A 2T8, QC Canada.}
\vspace{4pt}

\affiliation[2]{International Centre for Theoretical Sciences (ICTS-TIFR)\\
Shivakote, Hesaraghatta,
Bangalore 560089, India.}
\vspace{4pt}

\affiliation[3]{Tata Institute of Fundamental Research\\
Dr Homi Bhabha Road, Navy Nagar, Mumbai, 400005, India
\vspace{4pt}}

\affiliation[4]{Asia Pacific Center for Theoretical Physics \\
77 Cheongam-ro, Nam-gu, Pohang-si, Gyeongsangbuk-do, 37673, Korea.
\vspace{4pt}}

\begin{abstract}
{The principle of holography of information states that in massless gravity, it is possible to extract bulk information using asymptotic boundary operators. In our work, we study this principle in a linearized setting about empty flat space and formulate it using Dirac brackets between boundary Hamiltonian and bulk operators. We then address whether the storage of bulk information in flat space linearized massive gravity resembles that of massless gravity. For linearized massless gravity, using Dirac brackets, we recover the necessary criteria for the holography of information. In contrast, we show that the Dirac bracket of the relevant boundary observable with bulk operators vanishes for massive gravity. We use this important distinction to outline the canonical Hilbert space. This leads to split states, and consequently, one cannot use asymptotic boundary observables to extract bulk information in massive gravity. We also argue the split property directly without an explicit reference to the Hilbert space. The result reflects that we can construct \textit{local bulk operators} in massive gravity about the vacuum, which are obscured from boundary observables due to the lack of diffeomorphism invariance. Our analysis sheds some light on evaporating black holes in the context of the islands proposal.}
\end{abstract}

\maketitle

\section{Introduction}
The question of whether information about a bulk state can be extracted using boundary operators in a theory of gravity is an important one \cite{Laddha:2020kvp, Chowdhury:2021nxw, Raju:2020smc, Raju:2021lwh}. In this light, our motivation for this work is two-fold. Since a theory of gravity is a constrained system, it is only natural to ask whether such statements can be understood using the Dirac bracket formalism \cite{dirac, Henneaux:1992ig, Mukunda:1967xc}. The second goal is to use this formalism and apply it to Fierz Pauli massive gravity \cite{Fierz:1939ix}, which is an interesting modification to gravity and has been a subject of recent interest (see \cite{Higuchi:1986py, Porrati:2000cp, Kogan:2000vb, Arkani-Hamed:2002bjr, Creminelli:2005qk, Karch:2000ct, Porrati:2003sa, Goldhaber:2008xy, Aharony:2006hz, Hinterbichler:2011tt, deRham:2010ik, deRham:2010kj, Hassan:2011vm, Hassan:2011hr, Blake:2013owa, deRham:2014zqa, Capela:2012uk, Ghodrati:2017roz} for a sampling of works and references therein). Our chief motivation lies in the recent discussions of massive gravity in the context of islands for evaporating black holes \cite{Penington:2019npb, Almheiri:2019hni, Almheiri:2019psf, Penington:2019kki, Almheiri:2019qdq, Geng:2020fxl, Geng:2020qvw, Raju:2020smc, Ghodrati:2022hbb}. 

Whether information is holographically stored at the boundary is addressed by the following question: given access to asymptotic boundary operators, can we precisely determine the bulk state? This version of holography of information exists in massless gravity and is an essential consequence of the Gauss constraint. The crucial ingredient involved here is the boundary Hamiltonian, using which one can construct boundary operators that probe bulk physics \cite{Laddha:2020kvp, Chowdhury:2021nxw, Raju:2020smc, Raju:2021lwh, Chowdhury:2020hse, Banerjee:2016mhh, Raju:2018zpn, Chakraborty:2021rvy}. Related works discussing the localization of information in massless gravity are \cite{deMelloKoch:2022sul, Giddings:2021khn, Chowdhury:2022wcv, Bahiru:2022oas, Bahiru:2023zlc, Chakraborty:2023los, Chakraborty:2023yed}.

In massless gravity, the principle of holography of information implies that specifying a bulk state $\ket{\psi}$ outside a bounded region $B$ uniquely fixes it inside $B$ \cite{Chowdhury:2021nxw, Raju:2021lwh}. As a result, it is convenient to introduce \textit{split states}, i.e., states which can be arbitrary inside $B$ but are fixed on the complement of $B$. Generally, all field theories apart from massless gravity obey a \textit{split property} that the set of such states is non-empty \cite{Buchholz:1973bk}. However, the holography of information in massless gravity implies that the set of split states is empty. Keeping this in mind, in our work, we investigate the following objectives:
\begin{enumerate}
    \item To understand holography of information and split property using Dirac brackets by verifying known cases of linearized massless gravity and electrodynamics.
    \item To determine whether the property holds in massive gravity at a linearized level.
\end{enumerate} 
\subsection*{Brief description of results}

In our work, we account for constraints using Dirac brackets and use them to demonstrate the information stored at a linearized level for different constrained theories. Some related works on the phase space structure and the computation of Dirac brackets in massive gravity are given in \cite{Hinterbichler:2011tt, Kluson:2011qe, Jalali:2020dfj, Rodriguez-Tzompantzi:2022thd, Deser:2014vqa}. 

In \S \ref{app1}, we discuss physical observables in constrained theories and briefly review the Dirac bracket formalism for considering the constraints. Based on our discussion of physical observables, we develop a schematic argument for why we may be able to create \textit{local bulk operators} in massive gravity.  However, the presence of second-class constraints can render this picture wrong, and we need to verify the same by computing the Dirac bracket between the boundary Hamiltonian and an arbitrary bulk operator. We also argue that flat space massive gravity does not have asymptotic symmetries such as BMS supertranslations.

We argue that coupling linearized gravity (both massless and Fierz-Pauli massive gravity) to matter fields introduces inconsistencies in the structure of the constraints, including failure to close. As a demonstration, see Appendix \ref{appb} for the case of massless gravity and Appendix \ref{appc} for massive gravity, where in both cases, the constraints fail to close. Thus in principle, a complete calculation involves taking the full Einstein Hilbert action coupled to matter in the massless case. Similarly, we should couple matter covariantly to the full non-linear action for massive gravity \cite{deRham:2010ik, deRham:2010kj, Hassan:2011vm, Hassan:2011hr}.

In our work, we argue that the issues regarding split states can be understood even using a linearized analysis. We show that the Dirac matrix involving only the graviton phase space is sufficient to understand split states, while the remaining Poisson brackets are defined over the phase space of the complete gravity-matter theory. In other words, we demand that the brackets between constraints are computed only over the gravity phase space, which allows the constraints to close correctly. However, we do not put this restriction while computing the rest of the brackets, where the matter insertions are addressed adequately. Following this restriction, we also comment upon the vacua structure of massive gravity.

Intuitively this restriction is in line with our general expectation that the addition of matter should not drastically change the nature of the gravity constraint structure. Analogously in electrodynamics, the constraint analysis with or without including charged matter gives rise to the same Dirac matrix shown in Appendix \ref{app2}.

In \S \ref{sec3} and \S \ref{sec4}, we compute the Dirac matrix necessary to compute the brackets for massless and massive gravity, respectively. Using this, we address the extraction of bulk information using boundary operators for electrodynamics and massless/ massive gravity at leading order in perturbation theory. In \S \ref{sec5}, we define relevant boundary observables for massless and massive gravity and calculate the Dirac brackets. We also perform an alternate derivation of the Dirac brackets in Appendix \ref{appyc}.

For electrodynamics, using the Dirac matrix obtained in Appendix \ref{app2},  we find in \S \ref{app3} that one cannot use boundary operators to determine the bulk state, hence obeying the expected split property. For massless gravity, building upon the computation of the Dirac brackets in \S \ref{sec3} and \S \ref{sec5}, we obtain the necessary conditions for the lack of split states upon taking the Dirac bracket of boundary Hamiltonian with a bulk operator insertion in \S \ref{sec5.2}. 

For massive gravity, following \S \ref{sec4} and \S \ref{sec5}, we argue in \S \ref{kkek} that the computation of the Dirac bracket between the boundary operator with bulk operators vanishes. Upon quantization, lifting the Dirac bracket to the commutator between relevant operators acting on the Hilbert space implies that the commutator is zero. Due to this, in \S \ref{sec5.3}, we argue that in contrast to the principle of holography of information in massless gravity, we do not have an analogous statement in massive gravity. We argue this in two different ways: with and without an explicit reference to the Hilbert space of massive gravity. In \S \ref{discussion}, we discuss the potential limitations of our work, the implications of our results for evaporating black holes, and list some interesting directions.

\section{Physical observables and Dirac brackets}
\label{app1}
In gauge theory, there are different ways to address gauge redundancy. The general method to fix the redundancy is by defining gauge invariant observables. A useful subclass is to work with gauge invariant observables, which we can construct by fixing a good gauge choice, hence removing the redundancy (up to residual gauge, if any).

In massless gravity, there is a gauge redundancy, i.e., small diffeomorphisms, which die off at the asymptotic boundary of the spacetime. In flat space linearized gravity, these are
\beq \label{diffeo}
\delta h_{\mu \nu} = \del_{\mu}\zeta_{\nu}+\del_{\nu}\zeta_{\mu} + \text{O}\lc  \sqrt{G_N}\rc,
\eeq
where $\zeta$ parametrizes the diffeomorphisms at the linearized order. Hence the construction of physical observables in gravity is accomplished by demanding invariance under small diffeomorphisms characterized by \eqref{diffeo}, either by gauge fixing or by defining observables that commute with constraints. 
More generally, one cannot define local diffeomorphism invariant observables about \textit{simple} enough points in the phase space of massless gravity such as the one corresponding to the Minkowski vacuum, which corresponds to a maximally symmetric solution.\footnote{However one may be able to define diffeomorphism invariant observables for a complicated enough backgrounds by labelling the spacetime points using the values of curvature invariants \cite{DeWitt:1962cg, Bergmann:1960wb, Komar:1958ymq,Bahiru:2023zlc} as follows:
$$\phi(Z^a_0) = \int d^D x\, \phi(x) \delta(Z^a - Z_0^a) \rm {det} \frac{\partial Z}{\partial x}$$
where $Z^a$'s are curvature invariants. But this procedure fails to work for spacetimes which have a lot of symmetries e.g. Minkowski space. }

We can use gravitational dressing to construct observables invariant under small diffeomorphisms, where we dress bulk observables to the boundary.\footnote{See \cite{Donnelly:2015hta} for a detailed construction of gauge invariant operators using dressing in massless gravity and gauge theories.} In gauge theories, one can similarly construct similar gauge invariant observables, either by gauge fixing or by defining manifestly gauge invariant observables like Wilson loops.\footnote{Note that Wilson loops are not good observables in gravity beyond leading order, since the loops possess stress energy and hence backreact. However, they can undergo further gravitational dressing at subleading order and become good observables up to that order.}

The Fierz Pauli interaction term in massive gravity explicitly breaks the diffeomorphism invariance given in equation \eqref{diffeo}. Since small diffeomorphisms are no longer a symmetry of massive gravity, we need not define physical observables by methods such as dressing them using the boundary. 

In flat space massive gravity, since diffeomorphisms are no longer a symmetry, the subgroup of diffeomorphism group generating asymptotic symmetries such as supertranslations are absent.

From the phase space perspective, the fact that there is no gauge symmetry of the form \eqref{diffeo} for massive gravity is because the constraints of massive gravity are second-class and hence do not have any redundancy in the phase space. This feature contrasts the first-class constraints of massless gravity, which necessitate gauge fixing. Thus it naively seems that there can be \textit{local bulk observables} in massive gravity, which can completely hide from the boundary.

Despite the intuition from the gauge-fixing picture, we still need to consider the other second-class constraints for a consistent description. Due to these constraints, bulk observables may not be completely independent of the observables at the boundary. In this light, our work aims to understand whether these second-class constraints are sufficient for a boundary observer to fix the bulk state completely.

\subsection{Constraints and Dirac brackets} 
We will follow \cite{dirac} in our discussion. Here we will denote our set of constraints as $\{\Phi_i\}$. Given a Lagrangian $L$ for a constrained system, we have a set of primary constraints $\{\Phi^P_i\}$, which are independent relations between the fields $h$ and their canonical momenta $\Pi$. Let $H_0$ denote the Hamiltonian obtained by taking the Legendre transform of the Lagrangian $L$. We define the Dirac Hamiltonian to be
\beq \label{a1}
H = H_0 + v_i \Phi^P_i
\eeq
Recall that the Poisson bracket between two observables $F(x)$ and $G(y)$ is given by
 \beq
\{F(x), G(y)\} = \int d^{D-1}z \lc \frac{\delta F(x)}{\delta h(z)} \frac{\delta G(y)}{\delta \Pi(z)} - \frac{\delta G(y)}{\delta h(z)} \frac{\delta F(x)}{\delta \Pi(z)}\rc
\eeq
We first need to ensure whether the primary constraints are stable and use the stability to determine the parameters $v_i$ from \eqref{a1}. We check the stability by taking the Poisson brackets of primary constraints with the constrained Hamiltonian, i.e. $\{\Phi^P_i, H\}$, which either vanishes or gives us secondary constraints. Next, we need to check the stability of the secondary constraints, which may give us tertiary constraints. The process should be repeated for consistency of the constrained system until we have determined all possible constraints $\{\Phi_i\}$ and fixed the parameters $v_i$.

We can further classify the set of constraints $\{\Phi_i\}$ into two subsets: first-class and second-class. Second-class constraints are defined as constraints that do not commute among themselves i.e.
$$\{\Phi^s_i, \Phi^s_j\} \neq 0$$ on the constrained surface, while first-class constraints are defined as constraints that commute among themselves i.e.
$$\{\Phi^f_i, \Phi^f_j\} = 0$$
where we denote the first class constraints by $\Phi^f$, and the second class constraints by $\Phi^s$. 

The presence of first-class constraints in the system indicates the presence of gauge symmetry. Hence we need to fix a gauge corresponding to each of the first-class constraints. The set of first-class constraints $\{\Phi^f_i\}$ and the gauge conditions $\{G_i\}$ together form a system of second-class constraints. Once we obtain a system of second-class constraints, we can define the Dirac matrix as follows: 
\beq
C\lc \Phi_i, \Phi_j \rc \equiv \{ \Phi_i, \Phi_j \}.
\eeq
This matrix is now invertible since any constraint $\Phi_i$ gives a non-zero Poisson bracket with at least one other constraint\footnote{The first class constraints give non-zero Poisson brackets with the gauge constraints}. We then invert this matrix (not always), thereby obtaining the inverse $C^{-1}\lc \Phi_i, \Phi_j \rc$
 \beq
 C\lc \Phi_i, \Phi_k \rc \, C^{-1}\lc \Phi_k, \Phi_j \rc = \delta_{ij}
 \eeq
 With $C^{-1}_{ij} \equiv C^{-1}\lc \Phi_i, \Phi_j \rc$, the Dirac bracket between two observables $F(x_1)$ and $G(x_2)$ defined on the phase space is given by
 \beq \label{d.b.}
 \{ F(x_1), G(x_2) \}_{D.B.} = \{ F(x_1), G(x_2) \} - \int_{y_1} \int_{y_2}  \, \{ F(x_1), \Phi_{i}(y_1)\}\,  C^{-1}_{ij}\lc y_1, y_2\rc \, \{ \Phi_j(y_2), G(x_2)\}.
 \eeq 
 where we have used the notation $\int_{y_1} \int_{y_2} \equiv \int d^{D-1}y_1 \int d^{D-1}y_2$. Note that in \eqref{d.b.}, apart from the first term (i.e., the standard Poisson bracket), we also have the second term, which is the contribution due to the constraints.

\section{Linearized massless gravity: Dirac matrix}
\label{sec3}
Before addressing massive gravity, we will warm up with the Dirac matrix calculation for linearized massless gravity without matter, which will also help contrast results with the massive gravity calculation.

Let us begin with a convenient form for the action of the massless graviton:
 \begin{eqnarray}\label{gr0}
 L_{g} =  \frac{1}{\kappa^2}\left(-\frac{1}{2} \partial_\lambda h_{\mu\nu}\partial^\lambda h^{\mu\nu} + \partial_\mu h_{\nu\lambda}\partial^\nu h^{\mu\lambda}-\partial_\mu h^{\mu\nu}\partial_\nu h  +\frac{1}{2} \partial_\lambda h\partial^\lambda h \right) + \text{ boundary  terms}
 \end{eqnarray}
where the coefficient $\kappa^2$ is given by $\kappa^2 = 32 \pi G_N$, where $G_N$ is Newton's constant. The boundary terms in the Lagrangian \eqref{gr0} are chosen to simplify the momenta and the constraints, thereby giving us $\Pi_{00} = \Pi_{0i} = 0.$\footnote{Since we are working in asymptotically flat space, we can ignore possible boundary contributions to the pointwise constraints.} Using \eqref{gr0}, we compute the canonical momenta corresponding to $h_{\mu\nu}$:
 \beq\label{gmom}
\begin{split}
\Pi_{00} &= 0, \qquad \qquad \Pi_{0i} = 0\\
\Pi_{ij} &= \frac{\partial L}{\partial \dot h_{ij}} = \frac{1}{\kappa^2}\left(\dot h_{ij} - \dot h_{kk}\delta_{ij} -2 \partial_{(i}h_{j)0} + 2 \partial_{k}h_{0k}\delta_{ij}\right)
\end{split}
\eeq
The first line gives us $D$ primary constraints. Then the Hamiltonian for massless gravity is given by taking the Legendre transform of \eqref{gr0}:
\beq\label{3.1}
\begin{split}
H_0 &= \kappa^2\left(\frac{\Pi_{ij}^2}{2} - \frac{\Pi_{ii}^2}{2(D-2)}\right) +\frac{1}{\kappa^2}\left(\frac{1}{2}\del_{k}h_{ij}\del^{k}h^{ij} - \del_i h_{jk} \del^j h^{ik} + \del^{i}h_{ij}\del^j h_k^k -\frac{1}{2}\del_i h^j_j \del^i h^k_k \right)   \\
& \, - 2h_{0i} \del_j \Pi^{ij} - h_{00} \lc \nabla^2 h^i_i - \del^i \del^j h_{ij}\rc
\end{split}
\eeq
The constraints with Hamiltonian \eqref{3.1} are given by 
\begin{align} \label{fcon}
    \Pi^{00} & =0\\
    \Pi^{0i} & =0 \\
    \chi_0 = \{ \Pi^{00}, H_{\text{tot}} \} &= \nabla^2 h^i_i - \del^i \del^j h_{ij} \\
    \chi_i = \{ \Pi^{0i}, H_{\text{tot}} \}&=2 \del_j \Pi^{ij},
\end{align}
Since we have two primary constraints, the Dirac Hamiltonian is given by
\beq
H_{\text{tot}} = H_{0} + v_0\Pi^{00} +v_i \Pi^{0i}
\eeq
where $v_0$ and $v_i$ are undetermined constants that will be fixed. We can check that this system of constraints is first class since their Poisson brackets with themselves and the Hamiltonian vanish.
\subsubsection*{Gauge choice}
Given the above first-class constraints, we need to fix the redundancy in phase space. We do that by implementing constraints arising from fixing the gauge (i.e. small diffeomorphisms) and the undetermined constants $v_0$ and $v_i$ in the Hamiltonian.

A good gauge choice is fixing them so that the gauge constraints are orthogonal to the set of first-class constraints. Thus a natural guess for gauge conditions is the following\footnote{The numerical multiplicative factor in $K_0$ is chosen for later convenience.}:
\beq \label{3.7}
G_0 : h_{00} = 0, \qquad G_i : h_{0i} = 0, \qquad K_0: \frac{\Pi^k_{k}}{D-2} = 0 , \qquad \mathrm{and} \qquad K_j : \del^i h_{i j} = 0.
\eeq
In the rest of this section, we will use this choice to implement the Dirac procedure.

\subsubsection*{Dirac brackets}
Given the set of gauge conditions in \eqref{3.7}, we need to ensure their stability under time evolution, i.e., whether the above constraints give rise to new constraints after time evolution.
\beq \label{3.8}
\begin{split}
    \{ G_0 , H_{\rm tot}\} &= v_0\\
    \{ G_i , H_{\rm tot}\} &= v_i\\
    \{ K_0 , H_{\rm tot}\} &= - h_{00} \approx 0\\
    \{ K_j , H_{\rm tot}\} &=\del^i \Pi_{ij} - \frac{\del_{j} \Pi_k^k}{D-2} + 2\del_j \del^i h_{0i} \approx 0\\
\end{split}
\eeq
where $\approx$ denotes that the equation is valid on the constraint surface. From \eqref{3.8}, we see that a consistent choice of implementing the Dirac procedure is by setting $v_0 = v_i = 0$ since, in this case, we do not get any new constraints. From the perspective of counting degrees of freedom, we now have $4D$ second class constraints on an originally $D(D+1)$ dimensional phase space, thereby reducing the phase space dimensionality to $D(D-3)$. This reduction is consistent with the fact that the graviton has $\frac{D(D-3)}{2}$ degrees of freedom\footnote{The degrees of freedom in massless gravity in $D$-dimensions can be found out by counting the symmetric traceless representations of the little group $SO(D-2)$, giving rise to $\frac{D(D-3)}{2}$ polarizations of the standard graviton. Note here that $D\geq 3$.}. 

In hindsight, we will find that the above choice of gauge conditions is designed such that each gauge condition gives a non-zero commutator with exactly one of the first-class constraints, thereby helping us obtain a simpler yet non-singular Dirac matrix. Specifically, the non-zero elements of the constraint matrix are given by:
\beq
\begin{split}
    \{\Pi^{00} (x), G_0 (y)\} &=  - \delta^{D-1}(x-y)\\
    \{\Pi^{0i} (x), G_j (y)\} &= -\delta^i_{j}\delta^{D-1}(x-y)\\
    \{\chi_{0} (x), K_0 (y)\} &=  \nabla^2\delta^{D-1}(x-y)\\
    \{\chi_{i} (x), K_j (y)\} &= \left(\delta_{ij} \nabla^{2} + \pd_{i} \pd_{j} \right)\delta^{D-1}(x-y)\\
    \{K_{0} (x), K_i (y)\} &= \frac{1}{D-2}\pd_{i}\delta^{D-1}(x-y)
\end{split}
\eeq

For later convenience, we will rename the constraints as follows:
\beq\label{newl}
\begin{split}
&C_0 : \chi_{0}, \qquad   C_i : \chi_i , \qquad C_{D} : \Pi^{00} \qquad C_{D+i}: \Pi^{0i},\\
&C_{2D}: K_0, \qquad C_{2D + i}: K_i , \qquad  C_{3D}: G_0, \qquad  C_{3D + i}: G_i.
\end{split}
\eeq
In this new notation, we label the constraint matrix as $$C_{ab} = \{ C_a, C_b\}$$ where $a$ and $b$ run from $0 \cdots 4D-1$. Writing the matrix using the representation in the momentum space, we obtain the following:
\beq \label{3.11}
C(p) =\left[
 \begin{array}{cccccccc}
 0 & 0_{j} & 0 & 0_{j} & -\mathbf{p}^2 & 0_{j} & 0 & 0_{j} \\
 0^{i} & 0\indices{^{i}_{j}} & 0^{i} & 0\indices{^{i}_{j}} & 0^{i} & -(p^{i}p_{j}+\vb{p}^{2}\delta\indices{^{i}_{j}})& 0^{i} & 0\indices{^{i}_{j}} \\
 0 & 0_j & 0 & 0_{j} & 0 & 0_{j} & -1 & 0_{j} \\
 0^{i} & 0\indices{^{i}_{j}} & 0^{i} & 0\indices{^{i}_{j}} & 0^{i} & 0\indices{^{i}_{j}} & 0^{i} & -\delta\indices{^{i}_{j}} \\
\mathbf{p}^2& 0_{j} & 0 & 0_{j} & 0 & \frac{i p_{j}}{D-2} & 0 & 0_{j} \\
 0^{i} & (p^{i}p_{j}+\vb{p}^{2}\delta\indices{^{i}_{j}}) & 0^{i} & 0\indices{^{i}_{j}} & \frac{i p_{j}}{D-2} & 0\indices{^{i}_{j}} & 0^{i} & 0\indices{^{i}_{j}} \\
 0 & 0_{j} & 1 & 0_{j} & 0 & 0_{j} & 0 & 0_{j} \\
 0^{i} & 0\indices{^{i}_{j}} & 0^{i} & \delta\indices{^{i}_{j}} & 0^{i} & 0\indices{^{i}_{j}} & 0^{i} & 0\indices{^{i}_{j}} \\
\end{array}
\right] \ ,
\eeq
where we have used raised (lowered) indices on the matrix elements to abbreviate entries worth a column (row) array. Since the matrix given in \eqref{3.11} is non-singular, we can use it to compute the inverse matrix 

\beq \label{InvMlMat}
C^{-1}(p) =\frac{1}{2\vb{p}^{2}}\left[
 \begin{array}{cccccccc}
 0 & \frac{1}{D-2}\frac{i p_{j}}{\vb{p}^{2}} & 0 & 0_{j} & 2 & 0_{j} & 0 & 0_{j} \\
 \frac{1}{D-2}\frac{i p^{i}}{\vb{p}^{2}} & 0\indices{^{i}_{j}} & 0^{i} & 0\indices{^{i}_{j}} & 0^{i} & 2\delta\indices{^{i}_{j}}-\frac{p^{i}p_{j}}{\vb{p}^{2}}& 0^{i} & 0\indices{^{i}_{j}} \\
 0 & 0_j & 0 & 0_{j} & 0 & 0_{j} & 2\vb{p}^{2} & 0_{j} \\
 0^{i} & 0\indices{^{i}_{j}} & 0^{i} & 0\indices{^{i}_{j}} & 0^{i} & 0\indices{^{i}_{j}} & 0^{i} & 2 \vb{p}^{2}\delta\indices{^{i}_{j}} \\
-2 & 0_{j} & 0 & 0_{j} & 0 & 0_{j} & 0 & 0_{j} \\
 0^{i} & -2\delta\indices{^{i}_{j}}+\frac{p^{i}p_{j}}{\vb{p}^{2}} & 0^{i} & 0\indices{^{i}_{j}} & 0_{j} & 0\indices{^{i}_{j}} & 0^{i} & 0\indices{^{i}_{j}} \\
 0 & 0_{j} & -2\vb{p}^{2} & 0_{j} & 0 & 0_{j} & 0 & 0_{j} \\
 0^{i} & 0\indices{^{i}_{j}} & 0^{i} & -2\vb{p}^{2}\delta\indices{^{i}_{j}} & 0^{i} & 0\indices{^{i}_{j}} & 0^{i} & 0\indices{^{i}_{j}} \\
\end{array}
\right] \ .
\eeq

Notice that the inverse of the constraint matrix is non-local. Such non-localities are essential ingredients of a gauge invariant theory and encodes the structure of its Gauss law. We will later find that this feature gives rise to the property that the energy of field excitations within a spatial region is detectable from the boundary of the region. 

This concludes our analysis of the Dirac matrix for linearised gravity without matter.

\section{Linearized massive gravity: Dirac matrix}
\label{sec4}
We will now move on to computing the Dirac matrix for massive gravity without matter. The Fierz-Pauli action for a massive graviton is given by:
 \begin{eqnarray}\label{gr}
 L_{g} &=& \frac{1}{\kappa^2} \left(-\frac{1}{2} \partial_\lambda h_{\mu\nu}\partial^\lambda h^{\mu\nu} + \partial_\mu h_{\nu\lambda}\partial^\nu h^{\mu\lambda}-\partial_\mu h^{\mu\nu}\partial_\nu h  +\frac{1}{2} \partial_\lambda h\partial^\lambda h -\frac{1}{2} \mg^2(h_{\mu\nu}h^{\mu\nu}-h^2)\right)\nonumber\\
 &+& \text{ boundary terms}.
 \end{eqnarray}
 Again, as in the massless case, we have chosen boundary terms such that $\Pi_{00} = \Pi_{0i} = 0$ and $\kappa^2 = 32 \pi G_N$. In addition to the massless gravity Lagrangian, we now have the Fierz Pauli coupling term, with $\mg$ denoting the mass of the graviton. 
 
We can easily extend our analysis from the massless case to the massive case and similarly determine the remaining canonical momenta and the Hamiltonian. Since the kinetic part of the Lagrangian remains the same, the canonical momenta of the massive case are the same as for the massless case and are given by \eqref{gmom}. The massive gravity Hamiltonian is given by:
\beq
\begin{split}
H_g &= \kappa^2\left(\frac{\Pi_{ij}^2}{2} - \frac{\Pi_{ii}^2}{2(D-2)}\right) +\frac{1}{\kappa^2}\left( \frac{1}{2}\del_{k}h_{ij}\del^{k}h^{ij} - \del_i h_{jk} \del^j h^{ik} + \del^{i}h_{ij}\del^j h_k^k -\frac{1}{2}\del_i h^j_j \del^i h^k_k\right.    \\
& \left.\frac{1}{2}\mg^2 (h_{ij}h^{ij} - h^{i}_{i} h^{j}_{j}) - \mg^2 h_{0i}^2- h_{00} \lc \nabla^2 h^i_i - \del^i \del^j h_{ij} - \mg^2 h_k^k\rc \right)- 2h_{0i} \del_j \Pi^{ij} 
\end{split}
\eeq
As before, we again have two primary constraints, i.e. $\Pi_{00} = \Pi_{0i} = 0$. Using these primary constraints, the Dirac Hamiltonian is given by
\beq
H_{\text{tot}} = H_{g} + v_0\Pi^{00} +v_i \Pi^{0i}.
\eeq

\subsubsection*{Constraints and Dirac matrix}
As for the massless case, we systematically determine the constraints and repeat the Dirac procedure. Demanding stability of primary constraints under the action of the Hamiltonian, we find the following secondary constraints:
\beq
\begin{split}\label{cm1}
C_0 &= \{\Pi^{00}, H_{\rm tot}\} = (\nabla^{2}-\mg^{2})\grf_{j}^{j}-\pd_{i}\pd_{j}\grf^{ij}\\
C_i &= \{\Pi^{0i}, H_{\rm tot}\} = \pd^{j}\Pg_{ji}+\mg^{2}\grf_{0i} 
\end{split}
\eeq
Next, we demand the stability of these secondary constraints under the Hamiltonian and thereby obtain the following:
\beq\label{4.4}
\begin{split}
\Csn_{-1} &= \{C_0, H_{\rm tot}\}\approx   \mg^2\left(\frac{\Pg^{k}_{k}}{D-2} - \pd^{i}\grf_{i0}\right)  \\
\Csn_{-2} &= \{C_{-1}, H_{\rm tot}\}\approx  \mg^{4} \grf
\end{split}
\eeq
where $\approx$ denotes that the corresponding equation is valid on the constraint surface. The Poisson brackets of $C_{i}$ and $C_{-1}$ with the Hamiltonian can be set to zero by fixing the Lagrange multipliers $v_i$ and $v_0$, respectively. Thus, we have no tertiary constraints, and the system is consistent.

The above procedure leads us to a system of $2(\D+1)$ second class constraints provided we fix the Lagrange multipliers ($v_0$ and $v_i$) accompanying the primary constraints as follows
\begin{equation}
	H_{\tot} = H_g + \pd^{i}\grf_{i0} \Pg^{00} + \left(\pd^{j}\grf_{ji}-\pd_{i}\grf \right) \Pg^{i0}.
\end{equation}
In the limit $\mg \to 0$, \eqref{4.4} trivially vanishes and the constraints \eqref{cm1} ($\Csn_{a\ge 0}$ ) reduce to the massless gravity constraints \eqref{fcon}. Therefore the massive theory has two additional second-class constraints than the massless theory. As a cross-check, the above analysis leads to a correct determination of the degrees of freedom.\footnote{For theories with massive graviton, one needs to look at the symmetric traceless representation of the group $SO(D-1)$, which gives us $\frac{D^2 -D-2}{2}$ polarizations. This is valid for $D\geq 2$, and in particular, massive gravity in three dimensions has a propagating degree of freedom, whereas, in four dimensions, we have five polarizations.} 

Next, we define the constraint matrix
\begin{equation}
 C_{ab}(x,y) \equiv \bPB{\Csn_{a}(x)}{\Csn_{b}(y)} \end{equation}
where $a,\ b$ now spans $-2, -1, \dots, 2D-1$. We have defined $\{ C_{D}, C_{D+i} \} = \{\Pi^{00}, \Pi^{0i} \}$ and 
together with the constraints \eqref{cm1} and \eqref{4.4} they generate the constraint matrix
\begin{equation}
	C(x-y)=	 \mg^{2} \begin{bmatrix}
		0   & 	 \frac{d}{d-1}\mg^{4}     &    0    &  -  \mg^{2} \pd_{j}  &  -\mg^{2}   &  0_{j}   \\
		-\frac{d}{d-1}\mg^{4}   &  0   &    -  \nabla^{2} + \frac{d\mg^{2}}{d-1}   & 0_{j}   &     0    &    -  \pd_{j}  \\
		0 & \nabla^{2}- \frac{d\mg^{2}}{d-1}  &   0 &   \pd_{{j}} &  0 & 0_{j}   \\
		- \mg^{2}\pd^{i} &  0^{i} & \pd^{i} & 0\indices{^{i}_{j}}   &  0^{i} &   \delta\indices{^{i}_{j}}  \\ 
		\mg^{2}  &   0  & 0  & 0_{j}   &  0   & 0_{j}  \\
		0^{i}  & - \pd^{i}  &  0^{i}  & -\delta\indices{^{i}_{j}} & 0^{i}  &  0\indices{^{i}_{j}}
	\end{bmatrix} \delta^{d}(x-y) \, , \
\end{equation} 
where derivatives are taken with respect to the coordinate $x$ and $d=D-1$ denotes the dimension of the Cauchy slice. Here the raised and lowered indices denote rows and columns respectively. 

In the limit $\mg \to 0$, the above constraint matrix vanishes showing that the $2 \D$ constraints $\Csn_{a\ge 0}$ are first class in the massless limit. The remaining two constraints in \eqref{4.4} identically vanish. Note that the procedure to find the Dirac matrix in theory with first-class constraints requires gauge fixing as explained in \S \ref{sec3}. Hence the constraint matrix \eqref{3.11} for massless gravity cannot be directly obtained in the $\mg \rightarrow 0$ limit of the above matrix.

Fourier transforming $C(x-y)$, we get the momentum space constraint matrix
\begin{equation}
	C(p) = \mg^{2} \begin{bmatrix}
		0   & 	 \frac{d}{d-1}\mg^{4}     &    0    &  -  \mg^{2} i p_{j} &  -\mg^{2}   &  0_{j}   \\
		-\frac{d}{d-1}\mg^{4}   &  0   &     \vb{p}^2 + \frac{d\mg^{2}}{d-1}    & 0_{j}   &     0    &    - i p_{j}  \\
		0 & -\vb{p}^2 - \frac{d\mg^{2}}{d-1}   &   0 &  i p_{j} &  0 & 0_{j}   \\
		- \mg^{2} i p^{i} &  0^{i} & i p^{i} & 0 \indices{^{i} _{j}}   &  0^{i} &   \delta \indices{^{i}_{j}}  \\ 
		\mg^{2}  &   0  & 0  & 0_{j}   &  0   & 0_{j}  \\
		0^{i}  & - ip^{i}  &  0^{i}  & -\delta \indices{^{i}_{j}} & 0^{i}  &  0 \indices{^{i}_{j}}
	\end{bmatrix} \, ,
\end{equation} 
where we have used the momentum space representation of the delta function, $(2 \pi)^{d}\delta^{d}(x-y) = \int_{\vb{p}} e^{i \vb{p}.\vb{(x-y)}}$. The matrix $C(p)$ can be easily inverted to obtain the Dirac constraint matrix
\begin{equation}\label{inm}
C^{-1}(p) = \frac{1}{\mg^{4}} \frac{d-1}{d}   \begin{bmatrix}
                                0 & 0 & 0 & 0_{j}  & \frac{d}{d-1} & 0_{j} \\
	                            0 & 0 & -1 & 0_{j} & 0 &  -i p_{j}  \\
	                            0 & 1 & 0 &  ip_{j} & -\vb{p}^2 + \frac{d\mg^{2}}{d-1} & 0_{j} \\
	                            0^{i} & 0^{i} & ip^{i} & 0 \indices{^{i}_{j}} & 0^{i} & -p^{i}p_{j} - \frac{d\mg^{2}}{d-1}\delta \indices{^{i}_{j}}   \\
	                            -\frac{d}{d-1} & 0 & \vb{p}^2 - \frac{d\mg^{2}}{d-1} & 0_{j} & 0 & i p_{j}  \vb{p}^2 \\
	                            0^{i} & -i p^{i} & 0^{i} & p^{i}p_{j} + \frac{d\mg^{2}}{d-1}\delta\indices{^{i}_{j}} & i p^{i} \vb{p}^2 & 0\indices{^{i}_{j}}
                                  \end{bmatrix}.
\end{equation}
The main takeaway from the above analysis is that, unlike in the case of massless gravity, the Dirac matrix of a massive gravity theory has a local expression. This observation has important implications for the statement of holography of information. In particular, in \S \ref{kkek} we demonstrate that in contrast to the situation in massless gravity, massive gravity theories can hide information about bulk operator insertions from boundary operators.

\section{Boundary observables and Dirac brackets}
\label{sec5}
We will now utilize the Dirac matrices obtained for various theories, i.e. electrodynamics, massless gravity, and massive gravity, to calculate the Dirac brackets. 

\subsection{Boundary observables and Dirac brackets for massless gravity}

As in electrodynamics, the relevant boundary operator for massless gravity can be obtained from the Gauss constraint. The constraints for linearized massless gravity with matter are given in Appendix \ref{appb}, and from \eqref{gmat}, the Gauss constraint $\chi_0^m$ for massless gravity with matter insertion is given by
\beq \label{5.66}
\nabla^2 h^{k}_{k} - \del_i \del_j h^{ij} =- 16\pi G_N T_{00}.
\eeq
Given any bounded spatial region $V$, the Gauss constraint makes it possible to encode the energy of matter fields supported on it $\int_{V} T_{00}$ via an equivalent boundary operator given by:
\beq \label{5.77}
H_\del = \frac{1}{16\pi G_N} \int_{V} d^{D-1} x \, \del^i .(\del_j h_{i j}-\del_i h_k^k ) = \frac{1}{16\pi G_N}\int_{\del V} d^{D-2} x \, n^i .(\del_j h_{i j}-\del_i h_k^k  ) \ ,
\eeq
where $n$ denotes the unit normal to the boundary $\del V$ of $V$ (we review the analogous construction for electrodynamics in Appendix \ref{app2}). The Hamiltonian of the full theory can be obtained from $H_{\del}$ by taking the limit where $V$ includes the entire Cauchy slice containing it. 

Let us compute the Dirac bracket for the boundary operator $H_{\del}$ with some bulk matter insertion $O(z)$. Using the gravity constraints \eqref{newl}, since  $H_{\del}$ only depends on $h_{ij}$, we see that $H_{\del}$ has nonzero commutators only with the constraint $C_{2D} = K_0 $ given in \eqref{3.7}. The relevant commutator is given by:
\beq \label{eq:HC2D}
\begin{split}
 \{H_\del , C_{2D} (y) \} &=  \int  d^{d} z \, \frac{\del H_{\del}}{\del h_{ij}(z)} \frac{\del C_{2D}(y)}{\del \Pi^{ij}(z)}\\
 &= -\frac{1}{16\pi G_N} \int_{V} d^{d} x \   \nabla^{2} \delta^{d} (x-y)  =  \frac{1}{16\pi G_N} \int_{V} d^{d} x \int \frac{d^{d}p}{(2\pi)^{d}}  \,\mathbf{p}^2 e^{i \vb{p}. (\vb{x} -\vb{y})}  \ ,
 \end{split}
\eeq
where we have set $d=D-1$. The Dirac bracket of the boundary operator $H_\del$ with a bulk matter operator insertion $O(z)$ is given by:
\beq \label{5.8}
\begin{split}
\{H_\del , O(z) \}_{\text{D.B.}} &= -\int d^{d} y ~d^{d} z' \, \{H_{\del}, C_{a}(y)\}~C_{ab}^{-1}(y,z')~ \{C_b(z'), O (z)\}\\
&= \frac{1}{16\pi G_N}\int_{V} d^{d} x \int \frac{d^{d}p}{(2\pi)^{d}}\frac{d^{d}q}{(2\pi)^{d}} ~d^{d} y ~d^{d} z' \,  e^{i \vb{p}.(\vb{x} -\vb{y})} e^{i \vb{q}.(\vb{y}-\vb{z}')}\frac{\mathbf{p}^2}{\mathbf{q}^2}\{\chi_0^m(z'), O (z)\}\\
&=  \frac{1}{16\pi G_N} \int_{V} d^{d} x \ \{\chi_0^m(x), O (z)\} = \int_{V} d^{d} x \ \{T_{00}(x), O (z)\} \ ,
\end{split}
\eeq
where $\chi_0^m$ is the Hamiltonian constraint in the presence of matter, given in \eqref{gmat}. In the second step of \eqref{5.8}, we have used the fact that only the component $C^{-1}_{2D\,0}$ of the inverse constraint matrix contributes. Notice that in the limit where $V$ approaches the full spatial slice containing it, the right-hand side of \eqref{eq:HC2D} vanishes. However, the non-local factor arising from the inverse constraint matrix provides a measured counter-effect which makes \eqref{5.8} valid even for the full spatial slice. Thus the Dirac bracket of the boundary Hamiltonian with the observable $O(z)$ is equal to the Poisson bracket of the observable with the integrated Gauss constraint. This in turn, as expected, is equivalent to the Poisson bracket of $O(z)$ with the matter Hamiltonian.

\subsection{Boundary observables and Dirac brackets for massive gravity}
\label{kkek}

Like massless gravity, the relevant boundary Hamiltonian for massive gravity can be obtained from the Gauss constraint. From \eqref{b19} and \eqref{b20}, the Gauss constraint $\chi_0$ for massive gravity with matter insertion is given by
\beq
\nabla^2 h^{k}_{k} - \del_i \del_j h^{ij} = \mg^2 h^{k}_{k} - 16\pi G_N T_{00}.
\eeq
As in massless gravity, we can integrate the LHS of the Gauss constraint within a spacelike region $V$ and obtain the boundary operator $H_{\del}$, which is given by
\beq
H_\del = \frac{1}{16\pi G_N} \int_{V} d^{D-1} x \, \del^i .( \del_j h_{i j}-\del_i h_k^k )= \frac{1}{16\pi G_N} \int_{\del V} d^{D-2} x \, n^i .( \del_j h_{i j}-\del_i h_k^k).
\eeq
From the massive gravity constraints, we see that the boundary Hamiltonian fails to commute only with the constraint $C_{-1}$ (see Eqn \eqref{4.4}). The relevant commutator is given by:
\beq
\begin{split}
 \{H_\del , C_{-1} (y) \} &= \int d^{d} z \, \frac{\del H_{\del}}{\del h_{ij}(z)} \frac{\del C_{-1}(y)}{\del \Pi^{ij}(z)}\\
 &= -\frac{\mg^2}{16\pi G_N}  \int_{V} d^{d} x \,  \nabla^{2} \delta^{d} (x-y) =  \frac{\mg^2}{16\pi G_N}\int_{V} d^{d} x \int \frac{d^{d}p}{(2 \pi)^{d}} \,\mathbf{p}^2 e^{i \vb{p}. (\vb{x} - \vb{y})} \ ,
 \end{split}
\eeq
which is identical to \eqref{eq:HC2D} up to a factor of $\mg^{2}$. The Dirac bracket of the boundary operator with a bulk matter operator $O(z)$ is given by:
\beq \label{mgc}
\begin{split}
\{H_\del , O(z) \}_{\text{D.B.}} &=  -\int d^{d} y ~d^{d} z' \, \{H_{\del}, C_a(y)\}~C_{ab}^{-1}(y,z')~ \{C_b(z'), O (z)\}\\
&= \frac{d-1}{\mg^2 d} \frac{1}{16\pi G_N} \int_{V} d^{d} x \int \frac{d^{d}p}{(2\pi)^{d}} \frac{d^{d}q}{(2\pi)^{d}} d^dy\,d^{d}z' \,  e^{i \vb{p}.(\vb{x} -\vb{y})} e^{i \vb{q}.(\vb{y} -\vb{z}')}\\
& \qquad \qquad \qquad \qquad \qquad \qquad \mathbf{p}^2\big(\{C_0(z'), O (z)\}+  i q_i \{C_{D+ i}(z'), O (z)\} \big) \\
&=-\frac{d-1}{\mg^2 d} \frac{1}{16\pi G_N}\int_{V} d^{d}x \ \nabla^2\big(\{C_0(x), O (z)\} +  \del_i \{\Pi_{0 i}(x), O (z)\} \big)\\
&=-\frac{d-1}{\mg^2 d} \frac{1}{16\pi G_N}\int_{\del V} d^{d-1} x \ n^i\del_i\big(\{C_0(x), O (z)\}\big) \ .
\end{split}
\eeq
In the second equation above, we have identified the only contributing terms to be from $C^{-1}_{-1\, 0 }$ and $C^{-1}_{-1\, D+i}$. In the final step, we have neglected the contribution from the $C^{-1}_{-1\, D+i}$ terms as $O(z)$ is assumed to be a pure matter operator with trivial Poisson brackets with $\Pi_{0i}$. 

The result \eqref{mgc} differs from that of \eqref{5.8} fundamentally because, unlike in the case of massless gravity, the inverse constraint matrix contributing to \eqref{mgc} is local, thus allowing us to reduce the Dirac bracket to a pure boundary term. Therefore, when $O(z)$ is taken to be an operator insertion strictly in the bulk, we find that its Dirac bracket with the boundary operator $H_{\del}$ vanishes.

In Appendix \ref{appyc}, we treat the constraints of the free theory but substitute the equation of motion of $h_{0i}$ back into the action. We argue that at the classical level, the substitution makes sense. We use this to alternatively demonstrate that $\{H_\del , O(z) \}_{\text{D.B.}} = 0$.

\section{Vacua structure and split states} 

We will now utilize the Dirac brackets obtained for various theories, i.e. electrodynamics, massless gravity, and massive gravity, to investigate the existence of split states. In order to set up the stage for further discussions of constraints using Dirac brackets and their relation with split states, let us first begin with the case of electrodynamics. Readers familiar with split states in electrodynamics can directly skip ahead to \S \ref{2.1str}. 

\subsection{Split states in electrodynamics}
\label{app3}
We minimally couple a charged scalar $\phi$ to the electrodynamic field. The Hamiltonian for this system is given by
\beq \label{c0}
H_J = \int d^{d-1}x \lc -\frac{1}{2}\Pi^i\Pi_i + \frac{1}{4} F_{ij} F^{ij} -\del_i \Pi^i A_0 + \Pi_{\phi} \Pi_{\phi^*} - ie A_0 \lc \phi \Pi_{\phi} - \Pi_{\phi^*} \phi^*\rc +(D_i \phi)^* D_i \phi \rc \eeq
where $D_i \phi \equiv \del_i \phi + i e A_i \phi$ is the covariant derivative with respect to the gauge field, and $\Pi^i$ denotes the momentum conjugate to the electrodynamic field, while $\Pi_{\phi}$ denotes the momentum conjugate to the scalar field and is given by
\beq
\Pi_{\phi} = \dot{\phi^*} -ieA_0 \phi^*
\eeq
In terms of the gauge invariant fields, $\Pi^i = E^i$. Using \eqref{c0}, one can compute the Gauss constraint, which is given by:
$$\del_i\Pi^i - J^0 = 0$$
where $$J^0 = - ie \lc \phi \Pi_{\phi} - \Pi_{\phi^*} \phi^*\rc $$
is the matter current. As a consequence, the Gauss constraint implies that given a codimension one spacelike slicing $\Sigma$, measuring the integral of the electric field over the boundary gives us the total charge $Q = \int_{\Sigma} J^0$, i.e.
\beq \label{c1}
\int_{\Sigma} \del_i \Pi^i = \int_{\del \Sigma} n_i \Pi^i = Q.
\eeq
where $n_i$ is the outward pointing normal vector.

The boundary operator in \eqref{c1} is unique to electrodynamics due to the Gauss constraint and gives us the total charge. Now consider some matter insertion in bulk, denoted by the action of an operator $O(x)$. We want to determine whether the information content of the insertion can be obtained using a relevant boundary observable, i.e., the boundary operator defined in \eqref{c1}. 

The computation of the Dirac matrix for electrodynamics, both with and without matter, is performed in Appendix \ref{app2}. Using our analysis there, we are in a position to investigate the Dirac bracket of $\int_{\Sigma} \del_i \Pi^i $ with $O(x)$, which gives us
\beq \label{c2}
\begin{split}
     \Big\{ \int_{\Sigma} \del_i \Pi^i(x), O(z) \Big\}_{D.B.} &= \Big\{ \int_{\Sigma} \del_i \Pi^i(x), O(z) \Big\}\\ & -\int_{y_1} \int_{y_2} \, \int_{\Sigma} \del^2 \delta (x-y_1) \, \frac{1}{\del^2} \delta (y_1-y_2) \Big\{ \del_i \Pi^i(y_2) - J^0(y_2), O(z)\Big\}
\end{split}
\eeq 
 We will work with purely matter insertions $O(z)$ in the following discussion. Then the first Poisson bracket on the RHS of \eqref{c2} is zero, while the second bracket, upon integration by parts and using the Gauss law, takes the form
\beq \label{c3}
     \Big\{ \int_{\Sigma} \del_i \Pi^i(x), O(z) \Big\}_{D.B.}= -\int_{\Sigma} \Big\{ \del_i \Pi^i(x) - J^0(x), O(z)\Big\} = \{Q, O(z)\}
\eeq
where we have used $\{\del_i \Pi^i (x), O(z)\} = 0$. Thus we obtain an order one contribution due to the matter insertion. If we have a chargeless state, the Dirac bracket expectedly vanishes, whereas we obtain a finite contribution for the charged state. 

Lifting the Dirac brackets to operators acting on the Hilbert space, equation \eqref{c3} takes the following form \footnote{The lifting from phase space observables to operators acting on the Hilbert space is subject to the assumption that a suitable operator ordering prescription exists which removes any anomalous terms. This issue arises not only in electrodynamics but also in our later analysis of gravity, and we discuss more about this in \S \ref{discussion}.}
\beq \label{c333}
      \int_{\Sigma} \ls \del_i \Pi^i(x), O(z)  \rs = \ls Q, O(z)\rs
\eeq
Equation \eqref{c333} leads to the existence of split states in electrodynamics. To see this, note that the order one contribution on RHS is insufficient to specify the state of the bulk on the spacelike slicing. This is because the boundary operator $\int_{\Sigma} \del_i \Pi^i(x)$ can only measure the charge of the state. 

Consequently, there is an infinite degeneracy of states with a given electromagnetic charge, which all evaluate to the same value on the RHS of \eqref{c3}.  In this way, the Gauss constraint in electrodynamics cannot specify the state in question. Hence the split property holds since one cannot distinguish bulk states using the relevant boundary operator.

\subsection{Hilbert space and vacua structure in flat space gravity}
\label{2.1str}
In this subsection, we will revisit the canonical Hilbert space of asymptotically flat massless gravity and use it to construct the massive gravity Hilbert space analogously. The Hilbert space will be important in understanding the split property of flat space gravity in the later subsections. 

\subsubsection*{Massless gravity}
We begin by studying the vacua and boundary operators for flat space massless gravity \cite{PhysRevLett.46.573, Ashtekar:1981hw, Ashtekar:2018lor}. The asymptotic symmetries are implemented by subgroups of the diffeomorphism group, such as the BMS group, which generates supertranslations \cite{Bondi:1960jsa, Bondi:1962px, Sachs:1962zza, Sachs:1962wk, Strominger:2017zoo, Compere:2018aar}. 

Supertranslations (in $D=4$) are generated using supertranslation charges $Q_{lm}$ constructed by spherically smearing the Bondi mass aspect $m_B (u, \Omega)$ at $\mathcal{I^+_-}$:
\beq \label{2.1.1}
Q_{lm} = \frac{1}{4\pi G_N} \int \sqrt{\gamma} \, d^2\Omega \, m_B (u = -\infty, \Omega) \, Y_{l,m}(\Omega).
\eeq
We can separate $Q_{lm}$ into hard and soft components. Thus a specification of vacuum involves annihilation under positive modes of matter and gravity together with the eigenvalue under soft mode:
\beq \label{fsp1}
Q_{lm}\ket{0, \{s\}} = s_{l,m} \ket{0, \{s\}}
\eeq
Here in $\ket{0, \{s\}}$, the first label denotes the energy, while the second label specifies the supertranslation sector, i.e. the eigenvalue under the soft mode. Supertranslation sectors serve as different superselection sectors for the theory, and hence the flat space massless gravity Hilbert space is given by 
\beq \label{fsp3}
\mathcal{H} = \bigoplus_{\{ s \}} \mathcal{H}_{\{ s \}} 
\eeq
The ADM Hamiltonian $H_{\del}$ defined in \eqref{5.77} is simply $Q_{00}$ charge defined in \eqref{2.1.1} \cite{Arnowitt:1962hi, Regge:1974zd}. 

\begin{comment}
We now introduce an abstract projector onto the vacua subspace of the massless gravity theory \cite{Laddha:2020kvp,Banerjee:2016mhh}:
\beq \label{abspro}
\mathcal{P}_{0} = \lim_{a \to \infty} \exp\lc- a H_{\del} \rc.
\eeq
We can also define a more physical projector $\pdel$ which projects onto energies below an IR cutoff $\delta$ \cite{Chakraborty:2021rvy}:
\beq \label{abspro1}
\pdel = \Theta \lc \delta - H \rc.
\eeq
In \S \ref{sec5.2}, we will revisit the significance of $H_{\del}$ and the projectors by computing the Dirac bracket between $H_{\del}$ and a bulk matter insertion $O(z)$. Specifically, the projectors allow us to reconstruct bulk operators using the boundary.
\end{comment}
\subsubsection*{Massive gravity}

In order to construct the state space of linearized massive gravity, we note the following points:

\begin{enumerate}

    \item From our calculation in \eqref{mgc}, the Dirac bracket of $H_{\del}$ with a bulk matter insertion $O(z)$ is zero. Upon lifting operators to the phase space, we replace the Dirac bracket with commutators, thereby giving us
 \beq \label{six.12}
[H_{\del}, O(z)] = 0,
\eeq
where $H_{\del}$ is given by
\beq \label{six.13}
\quad H_{\del}  = \frac{1}{16\pi G_N} \int d^{D-2} x \, n^i .(\del^j h_{i j}-\del_i h_k^k ).
\eeq
    \item Since asymptotic symmetries are absent in massive gravity, in contrast to the massless theory, the vacua subspace of the flat space massive theory is labelled by a single sector rather than a direct sum over infinite supertranslation sectors as in \eqref{fsp3}. 
\end{enumerate}

Using these above points and from equations \eqref{six.12} and \eqref{six.13}, we can use the boundary operator $H_{\del}$ to label the states in the Hilbert space as $\ket{E,M}$ such that
\beq \label{fsp11}
H_{\del}\ket{E, M} = E \ket{E,M}. 
\eeq
Here the label $E$ denotes the eigenvalue under $H_{\del}$, and $M$ is a quantum number that labels the bulk matter and gravity insertions. In contrast to the massless case, the second label in \eqref{fsp11} does not denote the supertranslation sector but is closely related to the longitudinal mode of the massive graviton.

Due to the absence of asymptotic symmetries, we expect that the S-matrix defined over the state space $\{\ket{E, M}\}$ is infrared finite, i.e. as expected, there are no infrared divergences in pure massive gravity. 

\medskip
\begin{comment}
\textbf{Projectors onto the vacuum:} Using $H_{\del}$, we can again try to construct boundary projectors onto the vacua subspace using the boundary Hamiltonian $H_{\del}$, as in \eqref{abspro} and \eqref{abspro1}. The construction works within our linearized analysis since, by definition, the lowest value that $H_{\del}$ can take is zero. The boundedness of energy is crucial to defining the projectors at the linearized level.

However, from \S \ref{kkek}, the action of the projector $\mathcal{P}_{0}$ in \eqref{abspro} on a generic state does not project onto just the empty state $\ket{0,0}$, but projects onto a subspace of states labelled by $\ket{0,M}$.\footnote{This is in striking contrast to the massless gravity case where graviton and bulk matter insertions labelled by $M$ completely fix the boundary ADM energy $E$. We do not use $M$ as a separate quantum number in massless gravity since one label is enough to specify the state.} This is a significant departure from the flat space massless gravity case, and the above vacua structure will be important in addressing the questions regarding split states in massive gravity\footnote{At this stage, the fastidious reader may correctly anticipate that one can exploit the above-mentioned contrasting feature to construct split states in massive gravity.}.

However, defining projectors in this fashion makes sense only in linearized gravity. We comment on going beyond our linearized analysis in \S \ref{discussion}.
\end{comment}

\subsection{Split states in massless gravity}
\label{sec5.2}

Equation \eqref{5.8} states that the boundary Hamiltonian $H_{\del}$ knows about bulk insertions since the Poisson bracket of the matter insertion with the integrated stress energy component $T_{00}$ is the same as the commutator of the $H_{\del}$ with the matter insertion. This is in line with the Hamiltonian constraint analysis in equations (4.57) and (4.58) of \cite{Chowdhury:2021nxw}, where expanding the constraint at the second order in perturbation theory, $H_{\del}$ is related to the bulk energy. The bulk energy has a contribution from both gravity and matter, with the matter contribution being $T_{00}$. In our case, on the RHS of \eqref{5.8}, $O(z)$ clicks only with $T_{00}$ in the Poisson bracket. Thus the commutator of $H_{\del}$ with matter insertion gives us the same result as expected from the order-by-order expansion of the Hamiltonian constraint.

Given this consistency check, the statements about holography of information and split states follow as in \cite{Laddha:2020kvp}, which we briefly summarize. Using a Reeh Schlieder type argument, one can construct operators $Q_B$ supported near the boundary, which can act on a supertranslation vacuum $\ket{ 0, \{ s\}}$ to create any arbitrary state $\ket{ B, \{ s\}}$ in the supertranslation sector $\{ s\}$
\beq \label{qeye}
Q_B \ket{0, \{s\}} = \ket{ B, \{ s\}} + \text{O} \lc \sqrt{G_N} \rc.
\eeq
Henceforth we will ignore $\text{O} \lc \sqrt{G_N} \rc$ corrections. Thus using \eqref{qeye} any hard bulk operator\footnote{We only consider \text{hard} bulk operators here whose action does not change the superselection sector $\{ s\}$. } $O(z)$ can be written as
\beq \label{5.11}
O(z) = \sum_{mn} c_{mn} \ket{ m, \{ s\}} \bra{ n, \{ s\}} = \sum_{mn} c_{mn} \,  Q_m\ket{ 0, \{ s\}} \bra{ 0, \{ s\}} Q_n^{\dagger}.
\eeq
From equation \eqref{5.8}, we see that the matter insertion $O(z)$ leaves an imprint on the ADM energy since it does not commute with $H_{\del}$. Hence $H_{\del}$ can be labelled using the bulk matter-energy, which is positive definite in well-behaved theories.

Then away from our linearized case, in the full theory, one can use $H_{\del}$ to construct a boundary projector onto the vacuum of the theory on the lines of \cite{Laddha:2020kvp}. Using the Fock space representation of the projector we can write the operator representation in \eqref{5.11} as
\beq
O(z) = \sum_{mn} c_{mn} \,  Q_m P_0 \, Q_n^{\dagger}.
\eeq
which is completely expressed in boundary operators $Q_m$ and the projector $P_0$. Using the arguments of \cite{Raju:2020smc, Raju:2021lwh}, various statements regarding the holography of information follow from the above representation. Since bulk operators can be written as combinations of boundary operators, there cannot be split states since one can always utilize the boundary expression for the bulk operators to probe bulk physics. Thus the decomposition of the massless gravity Hilbert space $\mathcal{H}$ into 
\beq \label{hsf}
\mathcal{H} = \mathcal{H}_{A} \otimes \mathcal{H}_{\bar{A}} 
\eeq
is not allowed\footnote{The precise sense in which we refer to the factorization of Hilbert space is explained in \S \ref{sec5.3}.}, and correspondingly, the notion of split states in massless gravity does not exist \cite{Raju:2021lwh}. 

One can also use the more physical boundary projector $\pdel$ defined in \cite{Chakraborty:2021rvy} to express $O(z)$ in terms of boundary operators. However, this is a slightly more difficult task since this involves delicately tuning smearing functions on the complement of the bulk region we are interested in.

\section{Split states in massive gravity}
\label{sec5.3}
Since the Dirac bracket of $H_\del$ with bulk operator $O(z)$ is zero in massive gravity from equation \eqref{mgc}, we can hide any bulk matter insertion $O(z)$ from detection using the boundary operators. In this section, we will demonstrate the existence of split states in massive gravity in two ways: firstly by making an explicit reference to the Hilbert space outlined in \S \ref{2.1str}, and secondly by not making an explicit reference to the Hilbert space.

\subsection{Split states from the vacuum structure}
Following our notation introduced in \S \ref{2.1str}, from equation \eqref{mgc} we have 
\beq \label{mgcc}
[H_\del , O(z)] = 0.
\eeq
which implies that the we have simultaneous eigenstates $\ket{E, M}$. Thus there are arbitrarily many states with bulk matter insertions in the zero eigenvalue subspace of $H_\del$. 

This leads to split states since from equation \eqref{mgcc}, $H_{\del}$ or operators constructed using it can never detect matter eigenvalue $M$ in the bulk. In other words, one can \textit{shield} bulk excitations from any possible detection at the boundary. Hence, at our linearized analysis, there is no holography of bulk information at the boundary, which could be detected using boundary operators. Let us now precisely define what we mean by factorization of the Hilbert space.

 \subsubsection*{Approximate factorization of Hilbert space} The notion of Hilbert space factorization in eqn \eqref{hsf} is approximate. Even if we ignore any constraints, such a factorization is not allowed in a quantum field theory since the energy of such product states lies outside effective field theory. To work with such states within effective field theory, we should imagine some spatial separation between the regions $A$ and $\bar{A}$ larger than the UV length scale of effective field theory. %(a related aspect is that the algebra of observables in quantum field theory is Type III \cite{Witten:2021jzq}, which acts on a non-separable Hilbert space)%. 
 
Formally, introducing the spatial separation and momentarily ignoring the constraints, we can \textit{approximately factorize} the Hilbert space $\mathcal{H}$ upon spatially partitioning flat space into a bounded region $A$ and its complement $\bar{A}$ as follows
\beq \label{hsff}
\mathcal{H} = \mathcal{H}_{A} \otimes \mathcal{H}_{\bar{A}}.
\eeq
 Then the factorization in \eqref{hsff} implies that spatially partitioned split states of the following form exist in the Hilbert space $\mathcal{H}$:
\beq \label{hsff1}
\ket{\psi} = \ket{\psi_A} \otimes \ket{\psi_{\bar{A}}}.
\eeq
In our present case, region $A$ should be thought of as most of the bulk region while the complement region $\bar{A}$ is a small enough region sufficient to include the boundary. Then in massive gravity, states in the Hilbert space spanned by $\ket{\psi} =\ket{E,M}$ can be thought of as split states living in the above approximately factorised Hilbert space. 

This is roughly because we can modify the bulk state $\ket{\psi_A}$ in the region $A$ thereby changing the matter quantum number $M$, while preserving the boundary eigenvalue $E$ in the region $\bar{A}$ at the same time. Since there are no other relevant boundary operators that probe the bulk physics in the region $A$, changing $\ket{\psi_A}$ does not affect the state $\ket{\psi_{\bar{A}}}$ in region $\bar{A}$. Consequently, states of the form \eqref{hsff1} are allowed in the Hilbert space of massive gravity taking into account the constraints using Dirac brackets. 

\begin{comment}
Finally, in contrast to massless gravity, in electromagnetism, any boundary projector constructed using the boundary charge will project not only onto the vacuum but also to all states with zero charge.  Hence one cannot write a boundary expression for $O(z)$. Thus, the Hilbert space of massive gravity resembles that of electromagnetism since to fix the bulk state completely, we require an additional bulk label (for matter) than solely that of the boundary operator. Thus, it is no surprise that the split property is also violated in massive gravity.
\end{comment}

\subsection{Split states without the vacuum structure}

More generally, we do not need to reconstruct the Hilbert space in order to understand the split property. To see this, let us work with a non-gravitational theory first, and outline the split property.

Consider a density matrix $\rho$ defined on the spatially partitioned system acting on $\mathcal{H}$. We will be working with the algebra of observables $\mathcal{A}$ and $\mathcal{\bar{A}}$ acting on the previously-defined regions $A$ and $\bar{A}$ respectively. The algebras $\mathcal{A}$ and $\mathcal{\bar{A}}$ consist of products of bulk operators supported on $A$ and $\bar{A}$ respectively. 

We now look at operators $O_{A} \in \mathcal{A}$ and $O_{\bar{A}} \in \mathcal{\bar{A}}$, where elements from the algebras mutually commute, i.e., $\ls O_{A},  O_{\bar{A}}\rs =0$. Let us consider a bulk observable of the form $O = O_{A} O_{\bar{A}}$. The split property \cite{Haag:1992hx, fewster} implies that the expectation value of operator $O$ can be written as: 
\beq \label{hsff2}
\braket{O} \equiv \text{Tr}\lc \rho \, O \rc = \text{Tr}\lc \rho_A O_{A} \rc \text{Tr}\lc \rho_{\bar{A}} O_{\bar{A}} \rc
\eeq
where $\rho_A$ and $\rho_{\bar{A}}$ are density matrices such that they are unconstrained in the traced out regions $\bar{A}$ and $A$ respectively. Effectively the density matrix $\rho$ takes the following form
\beq \label{hsff3}
\rho  = \text{Tr}_{\bar{A}}\lc \rho_A  \rc \otimes \text{Tr}_A\lc \rho_{\bar{A}} \rc,
\eeq
which can be seen by substituting \eqref{hsff3} into \eqref{hsff2}.

\subsubsection*{Support of boundary operator $H_{\del}$}
In massless gravity, there are a couple of issues regarding the above construction. Firstly, from equation \eqref{5.8}, we have 
\beq
\ls H_{\del}, O_{A} \rs \neq 0.
\eeq
Hence the decomposition in \eqref{hsff2} does not work, since there always exists an operator $H_{\del}$ living in the region $\bar{A}$ which can be used to detect observables in $A$\footnote{From \eqref{5.8} since the commutator is non-zero, the ADM Hamiltonian $H_{\del}$, though supported in the region $\bar{A}$, does not belong solely to the algebra $\mathcal{\bar{A}}$.}. Correspondingly, the Hilbert space does not factorize, and the density matrix cannot be written in the form \eqref{hsff3} for massless gravity. 

The second issue arises since only small diffeomorphism invariant observables make sense. For instance, diffeomorphism invariant dressed operators obey $\ls O_{A},  O_{\bar{A}}\rs \neq 0$, and hence \eqref{hsff2} and \eqref{hsff3} do not hold. A more precise version involves defining the algebra of observables asymptotically. We refer the reader to \cite{Laddha:2020kvp, Raju:2021lwh} for a detailed treatment of this issue. 

In massive gravity, we do not need to work with asymptotic observables since diffeomorphism invariance is absent, and hence we can work with the previously defined algebra of observables. Since we do not need to dress the observables, we have $\ls O_{A},  O_{\bar{A}}\rs = 0$. Also for an observable $O_{A}$, the boundary operator $H_{\del}$ simply commutes with the spatially partitioned observables
\beq
\ls H_{\del}, O_{A} \rs = 0.
\eeq
As a consequence, equations \eqref{hsff2} and \eqref{hsff3} follow since we cannot use $H_{\del}$ which is supported in the region $\bar{A}$, to detect the state supported in the region $A$ any more. Thus from \eqref{hsff2} and \eqref{hsff3}, specifying information in the region $\bar{A}$ does not fix the state in the region $A$. Consequently, the existence of split states in massive gravity can be understood from lifting the Dirac brackets onto operator commutators acting on the Hilbert space. 

The appearance of split states is a significant departure from the case of massless gravity, where any matter insertion has a specific signature in correlators involving boundary operators. Massive gravity is not constrained enough to detect bulk insertions using boundary observables and their correlators. From our analysis in this subsection, it is clear that massive gravity resembles a non-gravitational QFT than massless gravity.

\section{Conclusion and discussion}
\label{discussion}

In our work, we have computed Dirac brackets in different settings and used them to investigate the issue of split states. More generally, we have addressed the following question regarding the principle of holography of information: given access to boundary operators, can one use them to identify a generic bulk state?  In linearized massive gravity, using our analysis of the Dirac brackets, it appears that such information is hidden from boundary operators, thereby allowing for split states.

We find that the Dirac bracket between the relevant boundary operator from the Gauss constraint and a generic bulk matter insertion is zero for massive gravity. This is consistent with our intuitively expected picture resulting from the lack of small diffeomorphisms in massive gravity. Thus one can create local bulk operators which can never be detected using the boundary operator $H_{\del}$ since the commutator is simply zero. This is a significant departure from massless gravity. We show that this leads to split states, and hence there is no holography of information in linearized massive gravity.

\subsubsection*{Potential limitations of our analysis}

Let us now discuss some potential limitations of our analysis:

\begin{enumerate}
    \item \textbf{Regarding the closure of constraints:} As discussed in Appendix \ref{applin}, from the perspective of constraints, one cannot consistently couple matter to the linearized gravity action since the constraint algebra does not close. The failure of linearised gravity-matter constraints to close introduces further constraints on the phase space. Introducing these additional constraints is inconsistent with the degree of freedom counting. In contrast, the case of electrodynamics is much simpler, where the Dirac matrix, upon the inclusion of charged matter, is the same as for the uncharged case (see Appendix \ref{app2}). The issues with consistently coupling matter with linearized gravity are an important feature contrary to our naive expectations. This issue is also demonstrated from the failure to impose $\del_{\mu} T^{\mu \nu} = 0$ in \cite{deRham:2014zqa}.\footnote{Very loosely, we can also argue that there should not be any corrections to the linearised Dirac matrix upon including matter since 
 naively coupling matter order by order in perturbation theory is 
 problematic, and as a consequence, the constraints 
 fail to close properly.}
    
    Motivated by electrodynamics, we circumvent this issue by taking the Dirac matrix of linearised gravity without matter to evaluate the Dirac brackets. This ensures that the algebra closes, and we use this matrix to compute Dirac brackets between observables, with subsequent Poisson brackets defined over the entire matter-gravity phase space. In other words, we restrict ourselves to the gravity phase space whenever we take the Poisson bracket between two different constraints but otherwise work in the entire matter-gravity phase space. A more satisfactory procedure would be to consider the full non-linear action coupled to matter \cite{deRham:2010ik, deRham:2010kj, Hassan:2011vm, Hassan:2011hr} and study the Dirac brackets, which is an open question. 

    \item \textbf{Holography of information:} In massless gravity, the principle of holography of information is a non-perturbative statement. Perturbatively we can demonstrate holography of information about low energy states. However, for heavy time-dependent states, as studied in \cite{Bahiru:2022oas, Bahiru:2023zlc}, within perturbation theory, it may be possible to construct operators which can commute with the Hamiltonian. We expect that there exist complicated observables using which we can probe the bulk, which incorporate non-perturbative effects.

    In our work, we restrict our analysis to linearized gravity over the empty flat background and restrict our statements to low energy states about the vacuum. At our present level of analysis, we have only checked in perturbation theory that the the commutator $[H_{\partial}, O(x)] =0$. This statement might receive non-zero corrections of O$(e^{-G_N^{-1}})$ but our present framework does not suffice to calculate such corrections. In particular, it is unclear whether the canonical vacuum in massive gravity satisfies necessary properties such as boundedness and whether one can define projectors onto the vacuum.  Thus we are unable to concretely establish whether holography of information in massive gravity is a non-perturbative statement or not.

    \item \textbf{Quantization ambiguities:} Generally, lifting constraints from the phase space to operator equations on the Hilbert space may introduce some corrections to the constraint algebra. For instance, we may have ambiguities proportional to $\delta (0)$ for first-class constraints. If such ambiguities arise, we need to implement a suitable choice of normal ordering that allows us to circumvent them.

    In our analysis of massive gravity, some simple operator-ordering ambiguities, such as ones resulting from the multiplication of canonical field with momenta, are absent since the constraints are all linear. There are seemingly no such obstructions to quantization at the level of our linearized analysis. 

    \item \textbf{Higgs-type mechanism and localization of information:} 
     A straightforward implication regarding the localization of information is as follows: the localization of information on the whole $AdS_5$ boundary is different from the Karch Randall type setups \cite{Karch:2000ct}, which have a massive graviton. In such setups, the massive graviton arises from a higher dimension. However, the crucial feature is that the boundary of the complete theory is not the same as the boundary of the dimensionally reduced theory, hence the difference in the localization of information. 

      Another question that may arise is how a possible Higgs mechanism which gives the graviton mass may change the localization of information. The naive picture is that breaking the diffeomorphism invariance by such a mechanism changes the localization of information, and consequently, even at the non-perturbative level, one may not be able to recover bulk information using just the boundary operators. Such issues still need to be better understood.

\end{enumerate}

\subsubsection*{Black hole evaporation}
We now briefly discuss some other implications of our work regarding black hole evaporation. Our analysis here indicates consistency with the arguments of \cite{Geng:2021hlu} that massive gravity at a linearized level allows for black hole evaporation using the islands formalism. This is because the Dirac bracket of operator insertions inside the disconnected entanglement wedge with the boundary Hamiltonian is zero, indicating consistency with the subregion duality.

However, since the Dirac bracket in massless gravity between the boundary Hamiltonian and the bulk Hamiltonian is not zero, operator insertions inside the entanglement wedge can potentially be detected using the boundary Hamiltonian. Whether our formalism sheds some light to circumvent this obstruction in massless gravity is an interesting question.

\subsubsection*{Other open questions}

We conclude our work with some other open questions. For massless gravity, we observe that the form of the constraint algebra of the linearised theory without matter and the complete non-linear theory with matter look similar, provided we fix the gauge appropriately. It would be interesting to investigate whether this observation also holds for the case of massive gravity. Stated differently, the question is whether we expect the form of the constraint algebra of non-linear massive gravity coupled to matter to be similar to the Fierz-Pauli case. Our analysis is plausibly valid for the full non-linear theory coupled to matter in such a case. A related point is whether our described vacua structure, which depends on our restrictive linearized analysis, generalizes to the non-linear case.

Given the constraint algebra of the non-linear action, an interesting question is whether a systematic procedure exists to truncate it to the constraint algebra from the quadratic action, i.e. to the linearized case. As we argued, to include matter, we need to consider the full non-linear Einstein action. However, is there any \textit{consistent} truncation of the full non-linear constraint algebra, which gives us the Dirac matrix of linearized gravity? 

 A slightly distant avenue is to understand whether there is any natural obstruction to lifting massive gravity phase space observables to state-independent operators \cite{Papadodimas:2015xma, Papadodimas:2015jra} and their subsequent dependence on late time effects \cite{Raju:2020smc, Chakravarty:2020wdm, Chakravarty:2021tia, Cotler:2016fpe}. We hope to address some of these issues in future work.

 \begin{acknowledgments}
We thank Suvrat Raju for suggesting the problem and for valuable suggestions regarding the work. We thank Claudia de Rham and Alok Laddha for useful correspondence, and to the anonymous referee for their comments in improving certain aspects of the draft. We are also grateful to Sayali Bhatkar, Simon Caron-Huot, Tuneer Chakraborty, Victor Godet, Priyadarshi Paul and Arnab Rudra for related useful discussions. We acknowledge the support of the Department of Atomic Energy, Government of India, under project number RTI4001. The work of J.C. is supported by Simons Collaboration on the Nonperturbative Bootstrap. The work of A.S. is supported by the JRG Program at the APCTP through the Science and Technology Promotion Fund and Lottery Fund of the Korean government, and the NRF grant from the Korean government (MSIT) (2022R1A2C1003182). A.S. is also supported by the Korean local governments of Gyeongsangbuk-do Province and Pohang City. A.S thanks KIAS for generous support during this work within the programs ``East Asian Joint Workshop 2022'' and ``Frontiers in Theoretical Physics 2022''.

\end{acknowledgments}

\appendix
 
\section{Dirac brackets for electrodynamics}
\label{app2}
We will first look at Dirac brackets for electrodynamics without matter and then derive the Dirac brackets after adding in the matter.

\subsection*{Electrodynamics without charges}
We will work with the Lagrangian given by
\beq \label{2.1}
L = -\frac{1}{4} F_{\mu \nu} F^{\mu \nu},
\eeq
where $F_{\mu \nu}$ denotes the field strength. Using this Lagrangian, we arrive at the primary constraint
\beq \label{2.2}
\Pi^0 = 0,
\eeq
where $\Pi^0$ denotes the standard canonical momentum. Then the Hamiltonian $H_0$ obtained from the Legendre transformation of \eqref{2.1} is given by
\beq
H_0 = \int d^{d}x \lc -\frac{1}{2}\Pi^i\Pi_i + \frac{1}{4} F_{ij} F^{ij} -\del_i \Pi^i A_0 \rc.
\eeq
To implement the Dirac bracket procedure, we will first add in the contribution from the primary constraint, i.e. 
\beq \label{2.4}
H = H_0 + v_0 \Pi^0,
\eeq
where our goal now is to fix $v_0$. The condition for the stability of the primary constraint gives us the Gauss constraint, which is a secondary constraint,
\beq \label{2.5}
\{ \Pi^0, H \} = \del_i \Pi^i. 
\eeq
We find that there are no further tertiary constraints because
\beq
\{ \del_i \Pi^i, H\} =0
\eeq
due to cancellations among terms resulting from integration by parts. Consequently, we have $v_0 =0$ in \eqref{2.4}, i.e. the constrained Hamiltonian is the same as obtained from Legendre transformation of the electrodynamics Lagrangian. Thus we have a system of first-class constraints characterized by the Hamiltonian $H$, and constraints \eqref{2.2} and \eqref{2.5}.

\subsubsection*{Gauge fixing}
Since we have a first-class system, we fix the gauge by putting in gauge conditions. A convenient choice is to choose a gauge that is orthogonal to the first-class constraints. In our case, this amounts to
\beq \label{2.7}
A_0 =0 \quad \text{and} \quad \del_i A^i =0.
\eeq
We rewrite the system of constraints given by \eqref{2.2}, \eqref{2.5} and \eqref{2.7} in the following ordered form
\beq \label{2.9}
\begin{split}
    C_0 &= \Pi^0(x)\\
    C_1 &= \del_i \Pi^i(x)\\
    C_{2} &= A_0 (x) \\
    C_{3} &= \del_i A^i (x) \\
\end{split}
\eeq
Using the above constraints, we have the following non-zero Dirac brackets
\beq \label{2.8}
\begin{split}
 \{ \Pi^0(x), A_0(y)\} &= -\delta^{d}(x-y)  \\
  \{\del_i \Pi^i(x), \del^j A_j(y)\} &= \nabla^2 \delta^{d}(x-y).
\end{split}
\eeq
Note that the plus sign in the expression for the second commutator in \eqref{2.8} arises due to the shifting of derivatives while performing integration by parts. Then the constraint matrix with the ordering in \eqref{2.9} is given by
\beq \label{2.10}
M(x-y) =\left(
 \begin{array}{cccc}
 0 & 0 & -1 & 0  \\
 0 & 0 & 0 & \nabla^2  \\
 1 & 0 & 0 & 0  \\
 0 & -\nabla^2 & 0 & 0  \\
\end{array}
\right)\delta^{d}(x-y)
\eeq
The inverse matrix of $M(x,y)$ from \eqref{2.10} is given by
\beq \label{2.11}
M^{-1}(x-y) =\left(
 \begin{array}{cccc}
 0 & 0 & 1 & 0  \\
 0 & 0 & 0 & -\frac{1}{\nabla^2}  \\
 -1 & 0 & 0 & 0  \\
 0 & \frac{1}{\nabla^2}  & 0 & 0  \\
\end{array}
\right)\delta^{d}(x-y)
\eeq

\subsection*{Electrodynamics with charges}
Recall from \eqref{c0} that the Hamiltonian for a charged scalar coupled to electrodynamics is given by
\beq
\begin{split}
H_J = &\int d^{d}x \lc -\frac{1}{2}\Pi^i\Pi_i + \frac{1}{4} F_{ij} F^{ij} -\del_i \Pi^i A_0 + \Pi_{\phi} \Pi_{\phi^*} - ie A_0 \lc \phi \Pi_{\phi} - \Pi_{\phi^*} \phi^*\rc + (D_i \phi)^* D_i \phi \rc
%& +\int d^{d-1}x \lc e^2 A_0^2 \abs{\phi}^2  \rc.
\end{split}
\eeq
Here we again have the primary constraint $\Pi_0 = 0$. As previously done, we write the constrained Hamiltonian as
\beq
H = H_J + v_0 \Pi^0
\eeq
The stability of the primary constraint gives us the secondary Gauss constraint, which is given by
\beq \label{b14}
\del_i \Pi^i - ie \lc \phi \Pi_{\phi} + \phi^* \Pi_{\phi^*} \rc  \, \equiv \, \del_i \Pi^i -J^0 =0. \footnote{As a comparison, for fermions, there is no electrodynamic contribution to the matter current.}
\eeq
Recall that now the Poisson bracket involves taking derivatives with respect to the scalar field and its conjugate momentum as well since a complete specification of the phase space involves accounting for the scalar field as well. We find that the secondary Gauss constraint is stable, i.e.
\beq
\{ \del_i \Pi^i-J^0, H\} =0
\eeq
due to cancellations among various terms, and use this to set $v_0 =0$, similar to the case of free electrodynamics. 
\subsubsection*{Gauge fixing and Dirac brackets}
Since the primary constraint remains unchanged and the secondary Gauss constraint receives an additive scalar contribution plus the contribution from $A_0$, a good orthogonal choice is to choose the same gauge conditions as previously chosen. This is because the extra charged contribution to the Gauss constraint commutes with the choice of gauge, which by construction gives a non-zero commutator with the free part. As a consequence, we have the constraints
\beq \label{b17}
\begin{split}
    C_0 &= \Pi^0(x)\\
    C_1 &= \del_i \Pi^i(x) - J^0\\
    C_{2} &= A_0 (x) \\
    C_{3} &= \del_i A^i (x) \\
\end{split}
\eeq
which gives us the exact same Dirac matrix as in \eqref{2.10} and its inverse in \eqref{2.11}.

\section{Constraints in linearized gravity with matter}
\label{applin}
In this Appendix, we covariantly couple matter to linearized massless and massive gravity. We show that the constraints do not close i.e., they become inconsistent with our expected counting of the degrees of freedom.

\subsection{Massless Gravity with minimally coupled matter}
\label{appb}
We will now minimally couple a scalar field to massless gravity using the stress-energy tensor. Using this, we will compute the constraints of this theory and calculate the Dirac bracket in this subsection. The action of minimally coupled matter to gravity is given by:
 \beq \label{scalar1}
  S_{\phi} = -\frac{1}{2}\int d^D x \, \sqrt{-g} (g^{\mu\nu}\partial_{\mu} \phi \partial_{\nu} \phi + \ms^2 \phi^2) 
 \eeq
 where $g$ is the determinant of the space-time metric and indices $\mu, \nu$ run from $0$ to $D-1$. Expanding the metric about the flat background (i.e. $g_{\mu \nu} = \eta_{\mu\nu} + h_{\mu \nu}$ and hence $g^{\mu \nu} = \eta^{\mu\nu} - h^{\mu \nu}$), at leading order, we obtain: 
 \beq \label{grsca}
  S_{\phi} =  \int d^D x \, \left(1 + \frac{h}{2}\right) \left(-\frac{\eta^{\mu\nu}}{2}\partial_{\mu} \phi \partial_\nu \phi - \frac{\ms^2}{2} \phi^2\right) + \frac{h^{\mu\nu}}{2}\partial_{\mu} \phi \partial_{\nu} \phi 
 \eeq
 where $h = Tr(h_{\mu\nu})$. In terms of the energy-momentum tensor $T_{\mu\nu}$, above action can be re-written as:
 \beq
  S_{\phi} = \int d^D x \, \lc -\frac{\sqrt{-g_b}}{2} h_{\mu\nu}T^{\mu\nu}(\phi, \dot \phi)  + \sqrt{-g_b} L_m (\phi, \dot \phi)  \rc
 \eeq
 where $L_m (\phi, \dot \phi)$ is the free-scalar Lagrangian, $g_b$ is the determinant of background metric (which is $\eta_{\mu\nu}$ in present case) and $T^{\mu\nu}$ is given by:
\beq
T^{\mu\nu} = \partial^\mu \phi \partial^\nu \phi - \frac{\eta^{\mu\nu}}{2}(\partial_\rho \phi \partial^\rho \phi + \ms^2 \phi^2)
\eeq
 The massless graviton Lagrangian $L_g$ is given in \eqref{gr0}. Hence the total action is given by:
 \beq \label{3.17}
 S = S_\phi + S_g
 \eeq
 where $S_g$ denotes the integral of $L_g$, with the $G_N$ dependence restored using an overall multiplicative factor $\kappa^2$.
 $$S_g = \frac{1}{\kappa^2}\int d^Dx \sqrt{-g} L_g$$
 \subsubsection*{Momenta and Hamiltonian}
 Using the combined action in \eqref{3.17}, we will now determine the canonical momenta and the Hamiltonian for our scalar-gravity theory. From the gravity part, we obtain the following expression for the momenta:
\beq \label{3.18}
\begin{split}
\Pi_{00} &= 0, \qquad \qquad \Pi_{0i} = 0\\
\Pi_{ij} &= \frac{\partial L}{\partial \dot h_{ij}} = \frac{1}{\kappa^2}\left(\dot h_{ij} - \dot h_{kk}\delta_{ij} -2 \partial_{(i}h_{j)0} + 2 \partial_{k}h_{0k}\delta_{ij}\right)
\end{split}
\eeq
From \eqref{3.18}, we find that we have two primary constraints $\Pi_{00}$ and $\Pi_{0i}$. The gravity Hamiltonian from \eqref{3.1} is given by:
\beq \label{3.19}
\begin{split}
H_g &= \kappa^2\left(\frac{\Pi_{ij}^2}{2} - \frac{\Pi_{ii}^2}{2(D-2)}\right) +\frac{1}{\kappa^2}\left( \frac{1}{2}\del_{k}h_{ij}\del^{k}h^{ij} - \del_i h_{jk} \del^j h^{ik} + \del^{i}h_{ij}\del^j h_k^k -\frac{1}{2}\del_i h^j_j \del^i h^k_k\right)\\
&- \frac{h_{00}}{\kappa^2} \lc \del_k^2 h^i_i - \del^i \del^j h_{ij} \rc- 2h_{0i} \del_j \Pi^{ij} 
\end{split}
\eeq
Next, we determine the canonical momenta of scalar field theory from the scalar Lagrangian given in \eqref{grsca},
\beq \label{3.20}
\pi_\phi = \frac{\partial L_\phi}{\partial \dot \phi} =  \dot\phi \left(1 + \frac{h}{2}\right) +  h^{00}\dot \phi + h^{0i} \partial_i \phi = \dot\phi \left(1 + \frac{h_{00} + h_{ii}}{2}\right)  + h^{0i} \partial_i \phi.
\eeq
We can invert \eqref{3.20} to determine $\dot \phi$ in terms of canonical variables, which will be useful to obtain the scalar Hamiltonian
\beq\label{phidot}
\dot \phi = \frac{\pi_\phi - h^{0i} \partial_i \phi}{ \left(1 + \frac{h_{00} + h_{ii}}{2}\right)}
\eeq
We can now obtain the Hamiltonian for the scalar field by Legendre transforming the scalar Lagrangian, i.e. $H_\phi = \pi_\phi\dot\phi - L_\phi$. Substituting equations \eqref{phidot} and \eqref{grsca} in the Legendre transform, we obtain:
\begin{eqnarray}\label{Hscalarfull}
H_\phi&=& \mathcal{E} - \frac{h_{00}}{2}\mathcal{E} + h_{0i} \ps \partial^i \phi
-\frac{h^k_k}{2}\left(\frac{\pi^2_{\phi}}{2} - \frac{(\nabla\phi)^2}{2}  - \frac{\ms^2 \phi^2}{2}\right) - \frac{1}{2}h_{ij}\del^i \phi\del^j \phi + O(h^2)
\end{eqnarray}
where the energy $\mathcal{E}$ is given by
\beq
\mathcal{E} = \frac{\pi^2_{\phi}}{2} + \frac{(\nabla\phi)^2}{2}  + \frac{\ms^2 \phi^2}{2}.
\eeq

Consequently, from equations \eqref{3.19} and \eqref{Hscalarfull}, and taking into account the primary constraints \eqref{3.18}, the full Hamiltonian is given by:
\beq
H_{\rm tot} = H_g + H_\phi + v_o \Pi^{00} + v_i \Pi^{0i}
\eeq
\subsection*{Closure of constraints and the Dirac matrix}
Let us now find the secondary constraints:
\beq
\begin{split}\label{gmat}
\chi^m_0 &= \left\{ \Pi_{00}, \int d^d x \, H_{\rm tot}\right\} =  \chi_0 + \frac{1}{2}\mathcal{E} \\
\chi^m_i &= \left\{ \Pi_{0i}, \int d^d x \, H_{\rm tot}\right\} = \chi_i + \ps \del_i \phi
\end{split}
\eeq
where $\chi_0$ and $\chi_i$ are the secondary constraints without matter. They are given by:
\beq
\chi_0 = \frac{1}{\kappa^2}\left(\del_i^2 h_k^k - \del_i\del_j h_{ij}\right)\qquad \qquad    \chi_i = -2 \del_j \Pi^{ij} 
\eeq
Next, we compute the tertiary constraints:
\beq
\begin{split}
\xi^m_0 &= \left\{ \chi^m_{0}, \int d^d x \, H_{\rm tot}\right\} =  \left\{\chi_0, \int d^d x \, H_{\rm tot}\right\}  + \frac{1}{2}\left\{\mathcal{E}, \int d^d x \, H_{\rm tot}\right\} \\
&= \xi_0  + \frac{1}{2}\left\{\mathcal{E}, \int d^d x \, H_{\phi}\right\} = -\del_i \del_j \Pi^{ij}+ \frac{1}{2}\del_i (\ps \del^i \phi) = -\del_i\chi_i^m \approx 0
\end{split}
\eeq

Next, we compute the commutator of $\chi_i^m$ with the Hamiltonian.
\beq
\begin{split}
\xi^m_i &= \left\{ \chi^m_{i}, \int d^d x \, H_{\rm tot}\right\} = \left\{\chi_i, \int d^d x \, H_{\rm tot}\right\}  +\left\{ \ps \del_i \phi, \int d^d x \, H_{\rm tot}\right\} \\
&=  \left\{\ps \del_i \phi, \int d^d x \, H_{\phi}\right\} = \ps\del_i \ps +\del_i \phi ( \del_k^2 \phi - \ms^2 \phi)
\end{split}
\eeq
At the perturbative level, it seems that the constraint algebra does not close. This can be seen from the fact that since $\xi^m_i$ is non-zero, its Poisson bracket with Hamiltonian we obtain a non-zero answer. Further it also seems that the constraints ($\chi_0^m$ and $\chi_i^m$) are no longer first class as well since $\{\chi_0^m(x), \chi_i^m(y)\} = \{T_{00}(x), T_{0i}(y)\} \neq 0$. 

However, this is a consequence of doing perturbation theory incorrectly. As an example, for the commutator $\{\chi_0^m(x), \chi_i^m(y)\}$ without matter, the gravity part of the constraints commute. However, if we keep the full non-linear correction in the gravity part of the Lagrangian, then the gravity part of the constraints does receive corrections, which then makes the constraints first class. 

\subsection{Minimally coupled matter to massive graviton}
\label{appc}
We again minimally couple the scalar field to gravity but with the Fierz Pauli action \eqref{gr}. As a consequence, the total action is given by:
 \beq
 S = S_\phi + S_g
 \eeq
where $S_g$ now denotes the Fierz Pauli massive gravity action.
\subsubsection*{Momenta and Hamiltonian}
In presence of matter, the full Hamiltonian is given by:
\beq \label{4.11}
H_{\rm tot} = H_g + H_\phi + v_o \Pi^{00} + v_i \Pi^{0i}
\eeq
where $H_\phi$ is given in \eqref{Hscalarfull}. The addition of matter does not affect the primary constraints, and consequently, they are still given by \eqref{gmom}. The stability of primary constraints leads to secondary constraints, which are given by
\beq
\begin{split}
\chi^m_0 &= \left\{ \Pi_{00}, \int d^d x \, H_{\rm tot}\right\} =  \chi_0 + \frac{1}{2}\mathcal{E} \\
\chi^m_i &= \left\{ \Pi_{0i}, \int d^d x \, H_{\rm tot}\right\} = \chi_i + \ps \del_i \phi
\end{split}
\eeq
where $\chi_0$ and $\chi_i$ denote secondary constraints without matter and are given by
\beq\label{b19}
    \chi_0 = \frac{1}{\kappa^2}\left((\del_i^2 - \mg^2)h_k^k - \del_i\del_j h_{ij}\right),\qquad
    \chi_i = -2\left(\del_j \Pi^{ij} + \frac{\mg^2}{\kappa^2} h_{0i}\right)
\eeq
Next, we demand the stability of secondary constraints, which give rise to the tertiary constraints:
\beq
\begin{split}\label{b20}
\xi^m_0 &= \left\{ \chi^m_{0}, \int d^d x \, H_{\rm tot}\right\} =  \left\{\chi_0, \int d^d x \, H_{\rm tot}\right\}  + \frac{1}{2}\left\{\mathcal{E}, \int d^d x \, H_{\rm tot}\right\} \\
&= \xi_0  + \frac{1}{2}\left\{\mathcal{E}, \int d^d x \, H_{\phi}\right\} = \xi_0 + \frac{1}{2}\del_i (\ps \del^i \phi) 
\end{split}
\eeq
where $\xi_0$ is again the constraint without matter and it is given by:
\beq
\xi_0 = -\del_i \del_j \Pi^{ij} + \frac{\mg^2}{D-2} \Pi^k_k - \frac{2 \mg^2}{\kappa^2}\del_i h_{i 0}.
\eeq
Hence the constraint $\xi_0^m$, is given by:
\beq
\begin{split}
\xi_0^m \equiv \xi' &=  -\del_i \del_j \Pi^{ij} + \frac{\mg^2}{D-2} \Pi^k_k - \frac{2 \mg^2}{\kappa^2}\del_i h_{i 0} +\frac{1}{2}\del_i (\ps \del^i \phi)\\
&= \frac{1}{2}\del_i \chi_i^m + \frac{\mg^2}{\kappa^2}\left(\kappa^2 \frac{\Pi^k_k}{D-2} - \del_i h_{0i}\right) \approx \mg^2\left( \frac{\Pi^k_k}{D-2} - \frac{\del_i h_{0i}}{\kappa^2}\right)
\end{split}
\eeq
Next, we compute the commutator of $\chi_i^m$ with the Hamiltonian \eqref{4.11}.
\beq
\begin{split}
\xi^m_i &= \left\{ \chi^m_{i}, \int d^d x \, H_{\rm tot}\right\} = \left\{\chi_i, \int d^d x \, H_{\rm tot}\right\}  +\left\{ \ps \del_i \phi, \int d^d x \, H_{\rm tot}\right\} \\
&= \frac{2 \mg^2}{\kappa^2} (\del^j h_{j i} - \del_i h - v_i ) + \left\{\ps \del_i \phi, \int d^d x \, H_{\phi}\right\} = \frac{2 \mg^2}{\kappa^2} (\del^j h_{j i} - \del_i h - v_i ) + \mathcal{B}_i
\end{split}
\eeq
where $\mathcal{B}_i = \left\{\ps \del_i \phi ,\int d^d x \, H_{\phi}\right\}$. We can solve for $v_i$ by demanding $\xi_i^m$ equals zero, which gives us
\beq
v_i = \del^j h_{j i} - \del_i h  + \frac{\kappa^2}{2 \mg^2}\mathcal{B}_i,
\eeq
and where upto the quadratic order in matter fields, $\mathcal{B}_i$ is given by:
\beq
\mathcal{B}_i =  \ps\del_i \ps +\del_i \phi ( \del_k^2 \phi - \ms^2 \phi)
\eeq
Next, we compute the quartic constraints by computing the commutator of $\xi_0^m$ with Hamiltonian \eqref{4.11}.
\beq
\begin{split}
\tilde \xi^m_0 &= \left\{ \xi^m_{0}, \int d^d x \, H_{\rm tot}\right\}=\frac{\mg^2}{D-2}\chi_0 + \frac{\mg^4}{\kappa^2}\frac{D-1}{D-2}h - \frac{1}{2}\del_i \mathcal{B}_i\\
&\approx \tilde\xi_0 - \frac{\mg^2}{2(D-2)}\mathcal{E}- \frac{1}{2}\del_i \mathcal{B}_i
\end{split}
\eeq
where $\tilde\xi_0$ is given by
$$\tilde\xi_0 = \frac{\mg^4}{\kappa^2}\frac{D-1}{D-2}h$$
Using the continuity equation, we have
\beq
\del^i B_i = \del^0 \del^i T_{0i}= \del_t^2 T_{00},
\eeq
Next, we find the Poisson bracket of the above constraints with total Hamiltonian:
\beq
\begin{split}
\left\{ \tilde \xi^m_{0}, \int d^d x \, H_{\rm tot}\right\} &=  \left\{\tilde \xi_0, \int d^d x \, H_{\rm tot}\right\}  - \frac{1}{2}\del_i \left\{\mathcal{B}_i, \int d^d x \, H_{\rm tot}\right\}+ \frac{\mg^2}{2(D-2)}\left\{\mathcal{E}, \int d^d x \, H_{\rm tot}\right\} 
\end{split}
\eeq
Again we can solve for $v_0$ by demanding the above equation to be zero. Various non-zero elements of the constraint matrix are given below:
\beq
\begin{split}
   \{\Pi_{00}(x), \tilde \xi^m_0 (y) \} &= \frac{\mg^4}{\kappa^2}\frac{D-1}{D-2} \, \delta(x-y), \qquad  \qquad  \{\Pi_{0i}(x), \chi^m_j (y)\} = \frac{2\mg^2}{\kappa^2}\delta_{ij}\,  \delta(x-y)\\
   \{\Pi_{0i}(x), \xi'(y) \} &= -\frac{ \mg^2}{\kappa^2}\del_{i}\delta(x-y), \quad
   \qquad \{\chi_0^m(x), \chi_i^m (y)\} = \frac{2 \mg^2}{\kappa^2}\del_i \delta(x-y) + \frac{1}{2} \mathcal{Q}_i\\
    \{ \chi^m_{0}(x), \xi'_{0} (y)\} &=  \frac{\mg^2}{\kappa^2} \ls \del_i^2 - \frac{D-1}{D-2} \mg^2\rs\delta(x-y)\\
     \{ \chi^m_{i}(x), \widetilde{\xi}^m_0 (y)\} &=  \frac{2\mg^4}{\kappa^2}\frac{D-1}{D-2} \del_i \delta(x-y)+ \frac{1}{2}\ls \del_t^2 + \frac{\mg^2}{D-2} \rs\mathcal{Q}_i\\
    \{ \xi'_0(x), \widetilde{\xi}^m_0 (y)\} &= -\frac{\mg^6}{\kappa^2} \lc \frac{D-1}{D-2}\rc^2 \delta(x-y), \qquad   \qquad
     \{ \chi^m_0(x), \chi^m_0 (y)\} =\frac{1}{4} \mathcal{P}\\ 
    \{ \chi^m_i(x), \chi_j^m (y)\} &=\mathcal{R}_{ij}, \qquad \qquad  \{ \widetilde{\xi}^m_0(x), \widetilde{\xi}^m_0 (y)\} =-  \frac{1}{4}\ls \del_t^2 + \frac{\mg^2}{D-2}\rs^2 \mathcal{P}\\ 
    \{ \chi^m_0(x), \widetilde{\xi}^m_0 (y)\} &= -\frac{1}{4} \ls \del_t^2 + \frac{\mg^2}{D-2} \rs \mathcal{P}\\ 
\end{split}
\eeq
where
\beq
\begin{split}
    \mathcal{P} &= \{T_{00}(x), T_{00}(y)\} = \lc \Pi_{\phi} (y) \frac{\del \phi (x)}{\del x^i} - \Pi_{\phi} (x) \frac{\del \phi (y)}{\del y^i} \rc \frac{\del}{\del x^i}\delta (x-y)\\
    \mathcal{Q}_i &= \{T_{00}(x), T_{0i}(y)\} = \lc \del_i \phi \del^k \phi \frac{\del}{\del x^k} - \pi_{\phi}^2 \frac{\del}{\del x^i}  + \ms^2 \phi \del_i \phi \rc\delta(x-y)\\
    \mathcal{R}_{ij} &=  \{T_{0i}(x), T_{0j}(y)\} = \lc \Pi_{\phi} (x) \frac{\del \phi (y)}{\del y^j} \frac{\del }{\del x^i} - \Pi_{\phi} (y) \frac{\del \phi (x)}{\del x^i} \frac{\del }{\del y^j} \rc \delta (x-y)
\end{split}
\eeq
By computing the inverse of the constraint matrix, we notice that we obtain the following type of inverse derivative dependence in Dirac brackets:
\beq \label{b32}
\frac{1}{\mg^2 - \del_i\del_j(\mathcal{R}_{ij}\mathcal{P}+ \mathcal{Q}_i\mathcal{Q}_j)}
\eeq
The above constraint algebra fails to close due to the presence of matter energy-momentum tensor on the RHS of the algebra. This can be seen by computing the Poisson bracket of RHS of any of the constraints above with the Hamiltonian. Since $\{\mathcal{P}, \mathcal{Q}_i\}$ (and other such combinations) is non-zero, the above algebra is not stable under Hamiltonian evolution.

This extra term in the denominator of \eqref{b32} is seemingly an artefact of perturbation theory. Since the constraints do not close now, they pose an inconsistency in the counting of the degrees of freedom. Consequently, we expect the extra term $\del_i\del_j(\mathcal{R}_{ij}\mathcal{P}+ \mathcal{Q}_i\mathcal{Q}_j)$ to go away as for the massless case upon the inclusion of higher order corrections \cite{deRham:2010ik, deRham:2010kj, Hassan:2011vm, Hassan:2011hr}. 

\section{Massive gravity constraints by substitution}
\label{appyc}

Let us look at a different way to compute Dirac brackets, where we substitute for some of the constraints. This procedure was used in \cite{Hinterbichler:2011tt}. Notice that the metric component $h_{0i}$ appears quadratically in the Fierz-Pauli lagrangian given in equation \eqref{gr}. Hence we can just solve for $h_{0i}$ using its equation of motion and substitute it back in the Lagrangian. This is the key difference from our previous treatment of constraints.The $h_{0i}$ EOM is given by:
\beq
h_{0i} = - \frac{1}{\mg^2}\del^j \Pi_{ji}
\eeq
This equation can also be obtained by setting the constraints $C_i$ (given in \eqref{cm1}) to zero. Now we can substitute $h_{0i}$ in the massive gravity lagrangian and then solve for constraints of the corresponding system. The Hamiltonian of this system is given by:
\beq
\begin{split}
H_g &= \kappa^2\left(\frac{\Pi_{ij}^2}{2} - \frac{\Pi_{ii}^2}{2(D-2)}\right) +\frac{1}{\kappa^2}\left( \frac{1}{2}\del_{k}h_{ij}\del^{k}h^{ij} - \del_i h_{jk} \del^j h^{ik} + \del^{i}h_{ij}\del^j h_k^k -\frac{1}{2}\del_i h^j_j \del^i h^k_k\right.    \\
& \left.\frac{1}{2}\mg^2 (h_{ij}h^{ij} - h_{kk}^2) - \mg^2 h_{0i}^2- h_{00} \lc \del_k^2 h^i_i - \del^i \del^j h_{ij} - \mg^2 h_k^k\rc \right)+ \frac{1}{\mg^2}( \del_j \Pi^{ij})^2 
\end{split}
\eeq
Since $\Pi^{00} = 0$\footnote{Since $h_{0i}$ is no longer a degree of freedom of the system, the corresponding momenta $\Pi^{0i}$ does not exist.}, this system has one secondary constraint given by:
\begin{eqnarray} 
    \Phi^g_1 &\equiv& ( \del_i \del^i - \mg^2) h_{j}^j - \del_i \del_j h^{ij}
\end{eqnarray}
One can readily compute the tertiary constraint :
 \beq \label{kekk1}
 \Phi^g_2 \equiv \{H_g, \Phi^g_1 \} = \frac{\mg^2}{D-2}\Pi_{ii} + \partial_i \partial_j \Pi_{ij}.
 \eeq
The stability of the above constraint under time evolution gives us the following further constraint:
\beq
 \Phi^g_3 \equiv \{H_g, \Phi^g_2 \} \approx - \frac{D-1}{D-2}\mg^2 h
\eeq
where $h = -h_{00} + h_{k}^{k}$. These set of constraints form a closed algebra. The constraint matrix is now a $4*4$ matrix whose various non-zero elements are given by
\begin{equation}
    \begin{split}
    \{ \Pi^{00}(p), \Phi^g_3(q) \}_{P.B.} &=\ -\mg^2\frac{D-1}{D-2}\delta^{D-1}(p-q),\\
     \{ \Phi^g_2(p), \Phi^g_1(q) \}_{P.B.} &=\ -\mg^4\frac{D-1}{D-2}\delta^{D-1}(p-q),\\
     \{ \Phi^g_1(p), \Phi^g_3(q) \}_{P.B.} &=\ 0, \\
     \{ \Phi^g_2(p), \Phi^g_3(q) \}_{P.B.} &=\ \mg^4 \left(\frac{D-1}{D-2}\right)\left( \mg^2\frac{D-1}{D-2} - p^2\right)\delta^{D-1}(p-q).
    \end{split}
\end{equation}
Hence the inverse of Dirac Matrix $C^{-1}(p)$ is given by:
\beq
C^{-1}(p) =\frac{1}{\mg^4}\frac{d-1}{d} \left(
\begin{array}{cccc}
 0 & \frac{d m^4}{d-1}-p^2 \mg^2 & 0 & \mg^2 \\
 p^2 \mg^2-\frac{d m^4}{d-1} & 0 & -1 & 0 \\
 0 & 1 & 0 & 0 \\
 -\mg^2 & 0 & 0 & 0 \\
\end{array}
\right)\delta^d(p-q)
\eeq
where $d = D-1$ is the dimension of the Cauchy slice. As expected, the above matrix is just a sub-matrix of \eqref{inm} (up to a factor of $\mg^2$ \footnote{This factor is different because of the difference in the definition of tertiary constraint.}).

We can now use this matrix to define the Dirac bracket. Since the inverse constraint matrix does not contain any derivatives in the denominator, using the analysis similar to the one discussed in \S \ref{kkek}, one can readily see that the Poisson bracket of boundary operator $H_\del$ with any bulk insertion $O(x)$ is zero. 

 \subsubsection*{Why substitution works classically?}
 \label{subs}
Roughly our action is of the form
\beq
L = L_0(x_i, \dot{x}_i) +\mg^2h_{0i}^2 + X h_{0i}
\eeq
such that $X$ is a function of $x_i, \dot{x}_i$. In the case of standard gravity with $\mg=0$, we have the constraint $X=0$, with $h_{0i}$ acting as a Lagrange multiplier. When $\mg^2 \neq 0$, we can rewrite the action as
\beq
L = L_0(x_i, \dot{x}_i) + m^2 \lc h_{0i} + \frac{X}{2\mg^2} \rc^2 -\frac{X^2}{4\mg^2}.
\eeq
Note that action is decomposed into a separate part for the $h_{0i}$ field, and a part containing $L_0(x_i, \dot{x}_i)$. Setting the positive definite second term in the above equation to zero gives us the equation of motion for $h_{0i}$. One can now always redefine the $h_{0i}$ field independently of $x_i, \dot{x}_i$
\beq
H_{0i} = h_{0i} + \frac{X}{2}
\eeq
such that there are no terms coupling the fields $x_i$ and $H_{0i}$. Thus we can independently minimize $H_{0i}$ without interfering with the variations of $L_0(x_i, \dot{x}_i)-\frac{X^2}{4\mg^2}$. Hence this substitution of the equation of motion is allowed. This is also confirmed by the counting of degrees of freedom. 

\textbf{Note:} This substitution is correct at a classical level but may pose some difficulties in quantum mechanics when we vary over the whole space of paths with weightage.

\bibliographystyle{JHEP}
\bibliography{refer1.bib}
\end{document}